%% file: phasepietadraftrev.tex
\documentclass[dvips,a4paper,12pt]{article}
\usepackage{graphicx}
\usepackage{amsmath}
\usepackage{authblk}
\usepackage[numbers,sort&compress]{natbib}
\usepackage[bookmarks=true,colorlinks=true,linkcolor=black,citecolor=magenta,breaklinks]{hyperref}
\usepackage{bm}
\usepackage{gensymb} %pour avoir le symbole \degree
\input{newcommands}

\newcommand{\unsurFQ}{{\frac{1}{F_\pi^4}}}

\newcommand{\sqsixfq}{{-\frac{\sqrt6}{F_\pi^4}}}  % -sign:  |1,1> = - |Eta Pi+> 

\newcommand{\fpiq}{F_\pi^4}

\newcommand{\LK}{L_K}
\newcommand{\Leta}{L_\eta}
\newcommand{\Lpi}{L_\pi}

\allowdisplaybreaks    % split equations across pages

\title{\bf\boldmath Form factors of the isovector scalar current and the $\eta\pi$
  scattering phase shifts      }

\author[a,b]{M. Albaladejo}

\affil[a]{\small Groupe de Physique Th\'eorique, IPN (UMR8608), Universit\'e
  Paris-Sud 11, Orsay}

\affil[b]{\small Instituto de F\'{\i}sica Corpuscular (IFIC), Centro Mixto
  CSIC-Universidad de Valencia}

\author[a]{B. Moussallam}

\begin{document}

\date{\today}

\maketitle

\begin{abstract}
A model for S-wave $\eta\pi$ scattering is proposed which could be realistic in
an energy range from threshold up to above one GeV, where inelasticity
is dominated by  the $K\bar{K}$ channel. 
The $T$-matrix, satisfying two-channel unitarity, is given in a form which
matches the chiral expansion results at order $p^4$ exactly for the
$\eta\pi\to\eta\pi$, $\eta\pi\to K\bar{K}$ amplitudes and approximately for
$K\bar{K}\to K\bar{K}$. It contains six phenomenological parameters.
Asymptotic conditions are imposed which ensure a minimal solution of the
Muskhelishvili-Omn\`es problem, thus allowing to compute the $\eta\pi$ and
$K\bar{K}$ form factor matrix elements of the $I=1$ scalar current from the
$T$-matrix.
The phenomenological parameters are determined such as to reproduce
the experimental properties of the $a_0(980)$, $a_0(1450)$ resonances, as
well as the chiral results of the $\eta\pi$ and $K\bar{K}$ scalar radii
which are predicted to be remarkably small at $O(p^4)$.
This $T$-matrix model could be used for a unified treatment of the
$\eta\pi$ final-state interaction problem in processes such as $\eta'\to
\eta \pi\pi$, $\phi\to\eta\pi\gamma$, or the $\eta\pi$ initial-state interaction
in $\eta\to3\pi$.
\end{abstract}

\tableofcontents

\section{Introduction}
The properties of the $\eta\pi$ scattering amplitude are much less known than
those of $\pi\pi$ or $ K\pi$ scattering. In the framework of three-flavour
chiral symmetry (in which the $\eta$ is a pseudo-Goldstone boson) a specific
prediction can be made that the $\eta\pi$ interaction should be considerably
weaker than the $\pi\pi$ or $ K\pi$ interactions~\cite{Bernard:1991xb} at low
energies. This feature has not yet been  verified either experimentally
or in lattice QCD. It is possibly related to the apparent absence of a broad
light $I=1$ scalar resonance.

A global description of $\pi\eta$ scattering (in particular of the elastic
channel and the leading inelastic channel $\pi\eta\to K\Kbar$) would enable
one to perform a universal treatment of the final-sate (or initial-sate)
interaction involving the $\pi\eta$ system. A particularly interesting
application would be to the $\eta\to3\pi$ amplitude. Precision
measurements of these decay modes should be exploited in an optimal way for the
determination of isospin violating quark mass ratios.  For this purpose, it is
necessary to combine chiral expansion expressions with general dispersive
treatments of rescattering~\cite{Kambor:1995yc,Anisovich:1996tx}.  An
extension of these approaches to include $\eta\pi$ rescattering would allow
one to take into account explicitly the $a_0-f_0$ ``mixing''
effect\footnote{This effect was first discussed in
ref.~\cite{Achasov:1979xc}. It can be seen as a superposition of the two
physical resonances $a_0(980)$, $f_0(980)$ in the $\eta\pi\to \pi\pi$
scattering amplitude.}, which was claimed to be
significant~\cite{AbdelRehim:2002an} for $\eta\to3\pi$.

The available experimental information on $\eta\pi$ scattering have
been derived via the final state interaction effects in production processes
and they concern, essentially, the properties of the
resonances. The two prominent resonances which have been observed in
the $S$-wave are the $a_0(980)$ and the $a_0(1450)$. We wish to
address here the problem of determining more global properties of the
$S$-wave amplitude i.e. the determination of phase shifts and
inelasticities in the small to medium energy range such as to be
compatible with the properties of the resonances and also obey
further theoretical constraints. 

Several models of the $\eta\pi$ $S$-wave scattering amplitude have
been proposed in the
literature~\cite{Oller:1998hw,GomezNicola:2001as,Black:1999dx,Furman:2002cg}.
Our approach enforces a correct matching with the chiral expansion of
the amplitudes at low energy in a way somewhat similar to
refs.~\cite{Oller:1998hw,GomezNicola:2001as}.  In addition, we propose
here to consider the form factor $F_S^{\eta\pi}$ (and $F_S^{K\Kbar}$)
associated with the scalar isovector current operator $\bar{u}d$, in
parallel with the $T$-matrix.
Form factors are the simplest quantities to which analyticity based
final-state interaction methods can be applied. We will follow  the same
general method which was proposed for the scalar isoscalar pion (and kaon) form
factors~\cite{Donoghue:1990xh} and proved capable of determining the scalar
radius of the pion $\braque{r^2}_S^{\pi\pi}$ rather accurately
(see refs.~\cite{Gasser:1983kx,Gasser:1990bv,Donoghue:1990xh,Moussallam:1999aq,Colangelo:2001df,Yndurain:2005cm,Oller:2007xd} for theoretical calculations, and refs.~\cite{Aoki:2009qn,Gulpers:2013uca,Aoki:2013ldr} for lattice determinations). Its application to the strangeness changing $K\pi$
scalar form factor and the corresponding scalar radius $\braque{r^2}_S^{K\pi}$
were discussed in refs.~\cite{Jamin:2001zq,Flynn:2007ki,ElBennich:2009da}. Form factors are constrained by
chiral symmetry at low energy and, even though the convergence of the three
flavour chiral expansion may be rather slow, one still expects correct order
of magnitudes to be provided at order $p^4$.  At this order, a simple relation
between the $\eta\pi$ and the $K\pi$ scalar radii is predicted
\be
\left. \dfrac{ \braque{r^2}_S^{\eta\pi}}
 {\braque{r^2}_S^{K\pi}}\right\vert_{p^4} = 0.52\pm0.02\ .
\en
This relation implies that the $\eta\pi$ radius is remarkably small
$\braque{r^2}_S^{\eta\pi}\simeq 0.1$ $\hbox{fm}^2$. We will show that this result
provides a stringent constraint in the determination of the phase shifts and
inelasticities.

The plan of the paper is as follows. We start with the chiral perturbation
theory (ChPT) expansions of the scalar form factors $F_S^{\eta\pi}$,
$F_S^{K\Kbar}$ and with the $\eta\pi$ and $K\Kbar$ scattering amplitudes at
next to leading order (NLO).  Next, we recall the general dispersive integral
equations from which one can compute the form factors starting from a given
$T$-matrix, provided suitable asymptotic conditions are imposed. We then
describe our chiral $K$-matrix type model for the $T$-matrix, which involves six
phenomenological parameters. It is designed such that, at low energies, the
contributions involving these parameters have chiral order $p^6$ (that is,
NNLO) and that a proper matching with the ChPT expressions at NLO holds
except, however, for the $K\Kbar\to K\Kbar$ amplitude, for which the matching
is only approximate. Finally, the determination of the phenomenological
parameters is discussed such as to satisfy the experimental constraints on the
$a_0$ resonances and the chiral constraints on the scalar form factors.

\section{\boldmath ChPT expansions of $\eta\pi^+$, 
$\bar{K}^0K^+$ form factors and scattering amplitudes} 
\subsection{Form factors and scalar radii}
Let us introduce the following two form factors associated with the isospin
one charged scalar operator $\bar{u}d$
\begin{align}\lbl{FFactdef}
B_0\,F_S^{\eta\pi}(s)= 
\braque{ \eta(p_1)\pi^+(p_2)\vert \bar{u}d(0) \vert 0}\nonumber\\
B_0\,F_S^{K\Kbar}(s) = 
\braque{ \kzerob(p_1)\kplus(p_2) \vert \bar{u}d(0) \vert 0}
\end{align}
where $s=(p_1+p_2)^2$. We have computed these form factors at next-to-leading
order (NLO) in the chiral expansion. The detailed expressions are given in
appendix~\sect{FFactorsp4}.
From eqs.~\rf{Fetapichir},~\rf{FKKchir} in that appendix, it is easy to derive
the expressions of the scalar radii, which are defined as
\be
\braque{r^2}_S^{PQ}= 6{\dot{F}_S^{PQ}(0)}/
{F_S^{PQ}(0)}\ .
\en 
For $\eta\pi$ and $K\Kbar$ one obtains
\begin{align}\lbl{r2etapip4}
& \braque{r^2}_S^{\eta\pi}=\dfrac{6}{\fpid}\Big[
            4\,L^r_{5}
+ \dfrac{1}{16\pi^2}\,\Big(-\dfrac{3}{4}L_K -\dfrac{11}{12}\Big)
+ \dfrac{\mpid}{3}\,\bar{J}'_{\pi\eta}(0)
          \Big]\\
& \braque{r^2}_S^{K\Kbar}= \dfrac{6}{\fpid}\Big[
4\,L^r_{5} + \dfrac{1}{16\pi^2}\Big(
-\dfrac{1}{2}\,L_\eta - \dfrac{1}{4}\,L_K 
-\dfrac{1}{2}\,R_{\pi\eta} -\dfrac{1}{4}\Big)
\ -\dfrac{2\mkd}{3}\,\bar{J}'_{\pi\eta}(0) \Big]\ ,
\end{align}
where $L_P$, $R_{PQ}$ are logarithmic functions of the pseudo-scalar
meson masses,
\be\lbl{logdefs}
L_P=\log\frac{m_P^2}{\mu^2},\quad
R_{PQ}=\frac{m_P^2\log(m_P^2/m_Q^2)}{m_P^2-m_Q^2}\ ,
\en 
with $\mu$ a renormalisation scale. These scalar radii depend on only one of
the Gasser-Leutwyler coupling constants~\cite{gl85}, $L^r_5$. It is
instructive to compare them with the analogous $K\pi$ scalar radius associated
with the strangeness changing scalar current, which also depends only on
$L^r_5$~\cite{gl85Kl3},
\be\lbl{r2Kpip4}
\braque{r^2}_S^{K\pi}= \dfrac{6}{\fpid}\Big[4\,L^r_{5}
-\dfrac{1}{8}\dfrac{1}{16\pi^2} \Big(6L_K +5R_{\pi K}+ R_{\eta K}
\Big) \Big] +\delta_2
\en
The explicit expression of $\delta_2$, from ref.~\cite{gl85Kl3}, is reproduced
in  appendix A. One remarks that the three scalar radii
$\braque{r^2}_S^{\eta\pi}$, $\braque{r^2}_S^{K\Kbar}$, $\braque{r^2}_S^{K\pi}$
have exactly the same dependence on the coupling $L^r_5$, which means that
they should be equal in the large $N_c$ limit of QCD. In reality, they are
rather different. Using e.g. $L^r_5=(1.23\pm0.06)\cdot10^{-3}$ (from
ref.~\cite{Bijnens:2014lea},  see sec.~\sect{ozidiscuss} below) one
finds\footnote{The following input numerical values are used
  throughout this paper (all in GeV): $m_\pi=0.139568$, $m_K=0.4957$,
  $m_\eta=0.547853$, $F_\pi=0.09221$.
} 
for $\eta\pi$ and $K\Kbar$
\be\lbl{r2Op4numvals}
\ba{ll}
\braque{r^2}^{\eta\pi}_S =& 0.092\pm0.007\ \hbox{fm}^2 \ ,\\
\braque{r^2}^{K\Kbar}_S  =& 0.136\pm0.007\ \hbox{fm}^2 \ ,\\
\ea\en
while for $K\pi$, one finds,
\be
\braque{r^2}^{K\pi}_S = 0.177 \pm0.007\ \hbox{fm}^2 \ .
\en
This shows that the $\eta\pi$ scalar radius is suppressed by a factor of two
as compared to the $K\pi$ scalar radius.
\subsection[Scattering amplitudes at $O(p^4)$]{\boldmath Scattering amplitudes at $O(p^4)$}
We consider the three scattering amplitudes involving the $\eta\pi^+$ and
the $\kzerob\kplus$ channels and we label the $\eta\pi^+$ channel as 1 and
the $\kzerob\kplus$ channel as 2.  At chiral order $p^2$ the
amplitudes read,
\be\ba{l@{}l}\lbl{Tstup2} 
{\cal T}^{11}_{(2)}(s,t,u)=& \dfrac{\mpid}{3\fpid}\\[0.3cm]
{\cal T}^{12}_{(2)}(s,t,u)=& \dfrac{\sqrt6}{12\fpid}\,(3s-4\mkd)\\[0.3cm]
{\cal T}^{22}_{(2)}(s,t,u)=& \dfrac{1}{4\fpid}(s+(t-u))\ .\\
\ea\en
The corrections of chiral order $p^4$ to these amplitudes can be
expressed in terms of a set of functions of one variable, analytic with a
right-hand cut, according to the so-called reconstruction
theorem~\cite{Stern:1993rg} (see also the
review~\cite{Zdrahal:2008bd}),
\be\ba{ll}\lbl{Decompth}
{\cal T}^{11}_{(4)}(s,t,u)
& =  U_0^{11}(s)+U_0^{11}(u)+W_0^{11}(t) \\[0.2cm]
{\cal T}^{12}_{(4)}(s,t,u)
& = U_0^{12}(s)+\left[ W_0^{12}(t)+(s-u)
  W_1(t) +(t\leftrightarrow u)\right]\\[0.2cm]
{\cal T}^{22}_{(4)}(s,t,u)
& = U_0^{22}(s)+(t-u)U_1(s)+V_0(t)+(s-u)V_1(t)+W_0^{22}(u)\ .\\
\ea\en
The detailed expressions of the functions $U_0^{ab}$, $W_0^{ab}$,
$U_j$, $V_j$ are given in appendix~\sect{Decompp4}. The resulting
amplitudes are equivalent to previous
calculations~\cite{Bernard:1991xb,GomezNicola:2001as}. We define the
partial-wave amplitudes as
\be\lbl{pwprojectdef}
T_J^{ab}(s)=\frac{1}{32\pi}\int_{-1}^1  
{\cal T}^{ab}(s,t(z^{ab}),u(z^{ab}))\,dz^{ab}
\en
such that the unitarity relation, in matrix form, reads
\be
\im\bm{T}_J(s) = 
\bm{T}_J(s)\,\bm{\Sigma}(s)
\, \bm{T}^\dagger_J(s)
=\bm{T}_J^\dagger(s)\,\bm{\Sigma}(s)
\, \bm{T}_J(s)
\en
with
\be\lbl{Sigdef1}
\bm{\Sigma}(s)=
\begin{pmatrix}
\sigma_1(s)\theta(s-(m_\eta+m_\pi)^2) & 0\\
0 & \sigma_2(s)\theta(s-4\mkd) \\
\end{pmatrix},
\en
and
\be\lbl{Sigdef2}
\sigma_1(s)=\frac{\sqrt{\lambda_{\eta\pi}(s)}}{s},\quad
\sigma_2(s)=\sqrt{\frac{s-4\mkd}{s} },\quad 
\lambda_{\eta\pi}(s)=(s-(m_\eta-m_\pi)^2)(s-(m_\eta+m_\pi)^2)\ .
\en
The relation between the partial wave $S$- and $T$-matrices then reads
\be\lbl{SJdef}
\bm{S}_J(s)=1 +2i \sqrt{\bm{\Sigma}(s)}\,
  \bm{T}_J(s)\,\sqrt{\bm{\Sigma}(s)}\ .
\en
In eq.~\rf{pwprojectdef}, $z^{ab}$ designate the cosines of the
centre-of-mass scattering angles, which are related to the Mandelstam
variables by
\be\ba{l}
t,u(z^{11})=\frac{1}{2}\left(  2\metad+2\mpid -s \pm
\dfrac{\lambda_{\eta\pi}(s) z^{11}-\Delta_{\eta\pi}^2}{s} \right)\\[0.2cm]
t,u(z^{12})=\frac{1}{2}\left( \metad+\mpid+2\mkd-s \pm 
\sqrt{\lambda_{\eta\pi}(s)}\,\sigma_2(s)\,z^{12}\right)\\[0.2cm]
t,u(z^{22})=\frac{1}{2}(4\mkd-s)(1\mp z^{22})\\
\ea\en
with $\Delta_{\eta\pi}=\metad-\mpid$. The first two of these relations
become singular when $s\to0$. This implies that the chiral expansions
of the $\eta\pi\to\eta\pi$ and $\eta\pi\to K\Kbar$ partial-wave
amplitudes become invalid when $s$ is too close to zero. If we assume
a domain of validity for the expansion of the unprojected amplitudes
when $|s|,\, |t|,\, |u| \lapprox 0.5$ $\hbox{GeV}^2$, then the chiral
expansions of the partial-wave amplitudes $T^{11}_J$, $T^{12}_J$
should converge with $s$ lying in the range $0.17 \lapprox s \lapprox
0.5$ $\hbox{GeV}^2$ and $0.05 \lapprox s \lapprox 0.5$ $\hbox{GeV}^2$
respectively.

From now on, we will consider only the $J=0$ partial-wave and will drop
the $J$ subscript. With the subscript now indicating the chiral order, the
$J=0$ partial-wave amplitudes at $O(p^2)$ are simply derived from~\rf{Tstup2}
\be\lbl{T0ijop2}
T^{11}_{(2)}(s)=\frac{1}{16\pi}\frac{\mpid}{3\fpid},\quad
T^{12}_{(2)}(s)=\frac{1}{16\pi}\frac{\sqrt6(3\,s-4\,\mkd)}{12\fpid},\quad
T^{22}_{(2)}(s)=\frac{1}{16\pi}\frac{s}{4\fpid}\ .
\en
The corrections of chiral order $p^4$  to these $J=0$ partial-wave amplitudes
can be written as
\be\lbl{T4leftright}
T^{ij}_{(4)}(s)=\dfrac{1}{16\pi}\left( U_0^{ij}(s) +\hat{U}_0^{ij}(s)\right)
\en
where
\begin{align}\lbl{U0hatdef}
& \hat{U}_0^{11}(s)=\frac{1}{2}\int_{-1}^1 dz^{11}\,
\left( U_0^{11}(u) + W_0^{11}(t)\right)\nonumber \\
& \hat{U}_0^{12}(s)= \int_{-1}^1 dz^{12}\,
\left( W_0^{12}(t)+(s-u)\,W_1(t) \right)\\
& \hat{U}_0^{22}(s)= \frac{1}{2}\int_{-1}^1 dz^{22}\,
\left( V_0(t)+ (s-u)\,V_1(t) +W_0^{22}(u)\right) \nonumber 
\end{align}
The functions $\hat{U}_0^{ij}(s)$ carry the left-hand cuts of the partial-wave
amplitudes $T^{ij}$. These cuts  are as follows~\cite{Kennedy1961}:
\begin{itemize}
\item[$T^{11}$:] A real cut on $[-\infty,(m_\eta-m_\pi)^2]$ and a complex
  circular cut centred at $s=0$ with radius $\Delta_{\eta\pi}$.
\item[$T^{12}$:] A real cut on $[-\infty,0]$ and a complex
  quasi-circular cut which intersects the real axis at
  $-\Delta_{\eta\pi} m_K/(m_K+m_\eta)$ and  $\Delta_{\eta\pi}
  m_K/(m_K+m_\pi)$. 
\item[$T^{22}$:] A real cut on $[-\infty,4\mkd-4\mpid]$.
\end{itemize}
As a final remark, at NLO, each one of the functions $U_0^{ij}$, $W_0^{ij}$,
$U_1$, $V_j$ can be written as the sum of a polynomial part and one involving
a combination of functions $\bar{J}_{PQ}$ (see appendix~\sect{Decompp4}). The
latter part is constrained by unitarity. For instance, for the functions
$U_0^{ij}$, one can write, in matrix form,
\be\lbl{U0decomp}
\bm{U_0}(s)= \bm{P_0}(s)+ (16\pi)^2 \,\bm{T}_{(2)}(s)
\begin{pmatrix}
\bar{J}_{\pi\eta}(s) & 0 \\
0 & \bar{J}_{K \Kbar}(s) \\
\end{pmatrix} \bm{T}_{(2)}(s)\ .
\en

\subsection[Influence of the  $1/N_c$ suppressed couplings]{\boldmath Influence of the  $1/N_c$ suppressed couplings}
\begin{figure}
\centering
\includegraphics[width=0.499\linewidth]{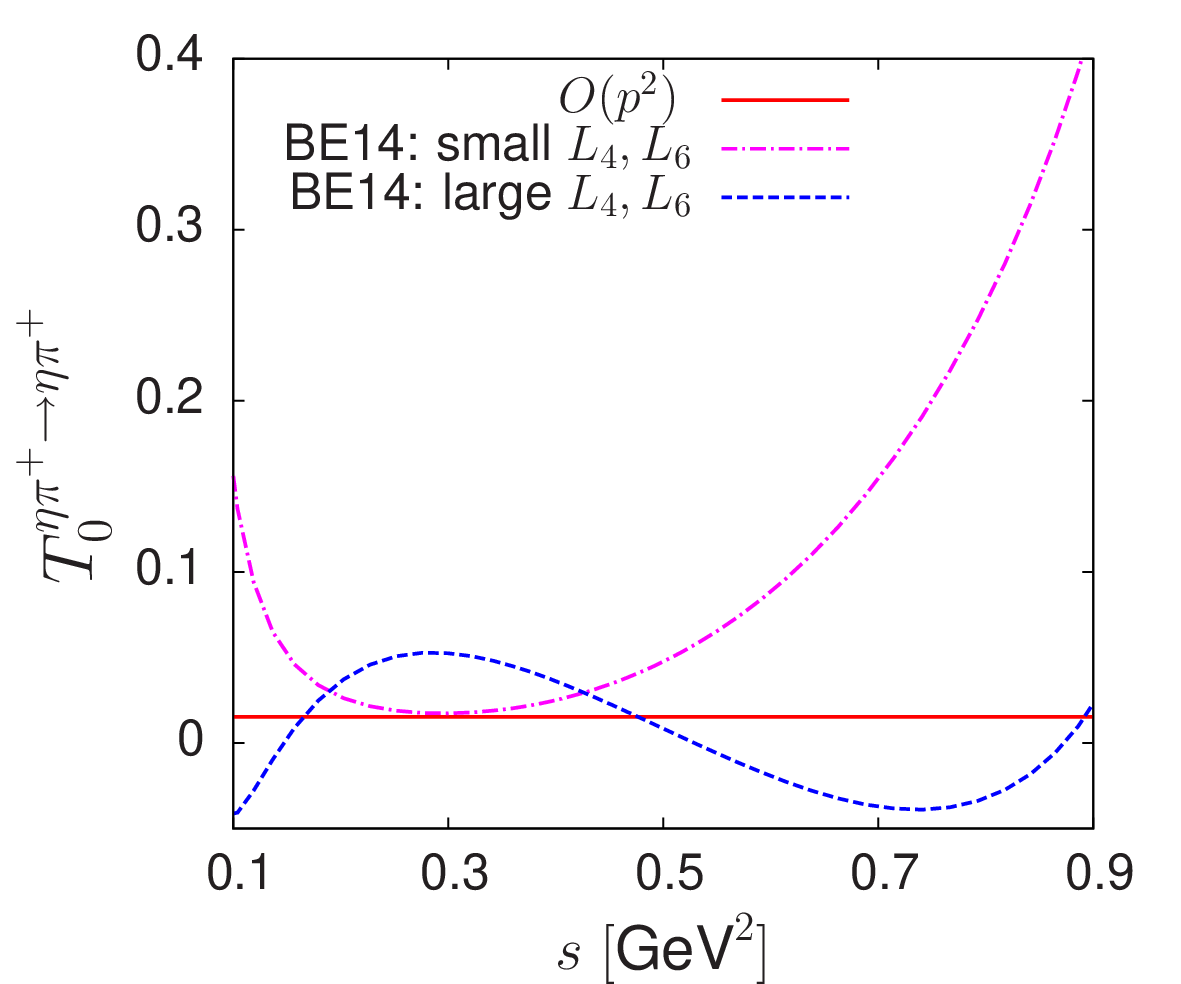}\includegraphics[width=0.499\linewidth]{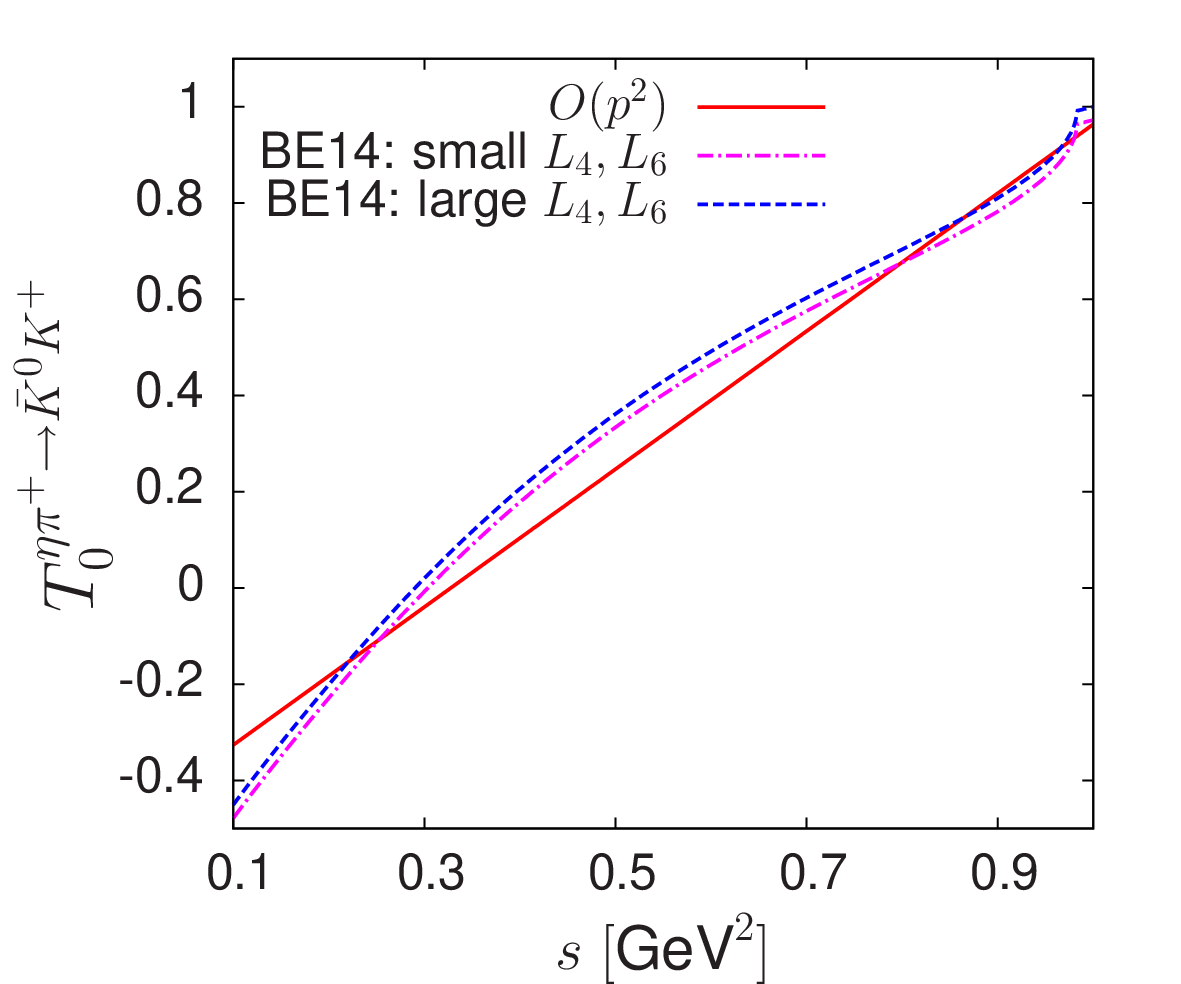}\\
\includegraphics[width=0.5\linewidth]{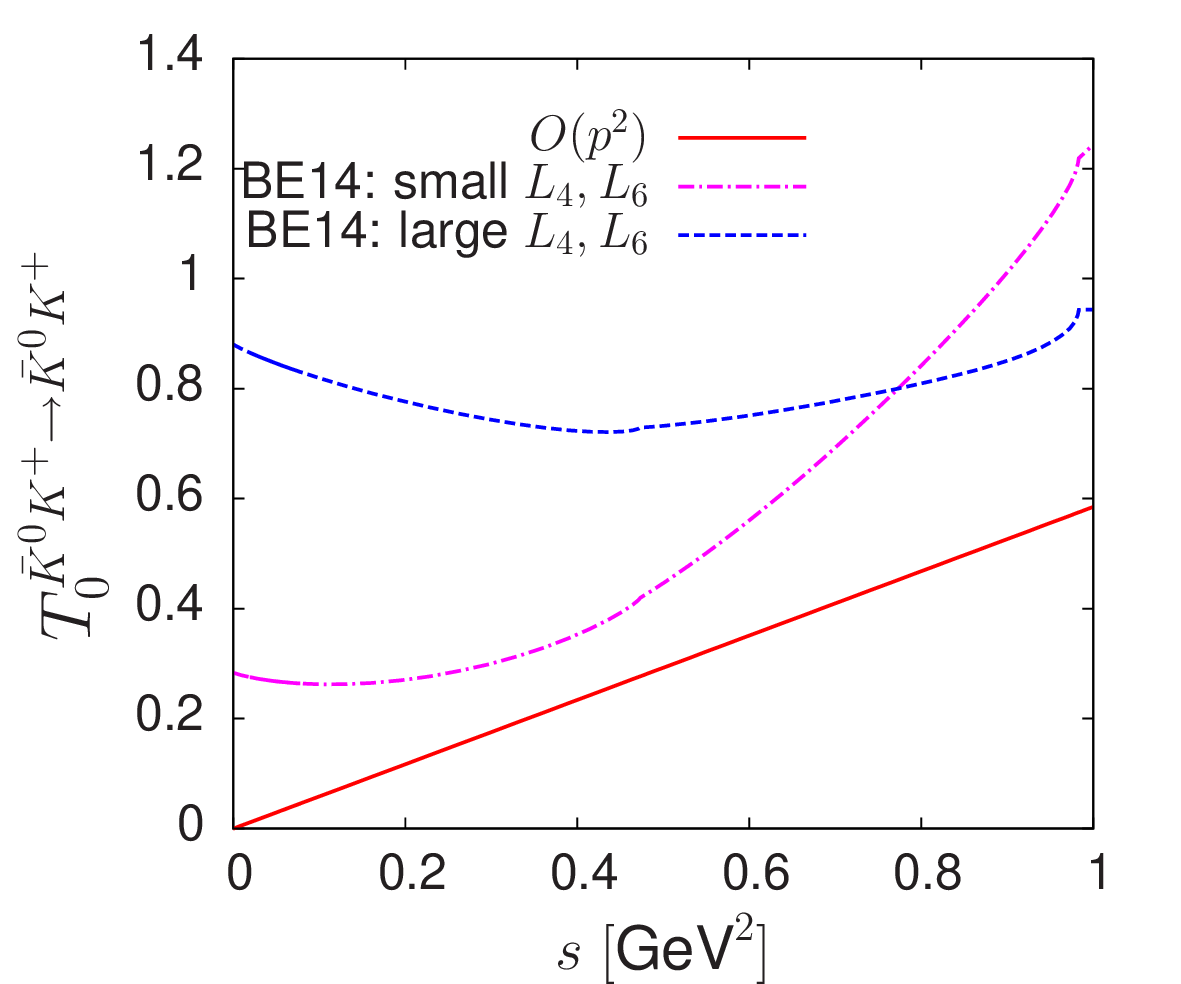}\\
\caption{\small Real parts of the three $J=0$ partial-wave amplitudes 
$\eta\pi^+\to \eta\pi^+$, $\eta\pi^+\to \kzerob\kplus$ and
  $\kzerob\kplus\to \kzerob\kplus$ at leading and next-to-leading
  order in ChPT.} 
\label{fig:AmplitChir}
\end{figure}

\lblsec{ozidiscuss}
%%%%%
\begin{table}[h]
\centering
\bt{ccccccccc}\hline
     &$10^3\,L_1^r$ &$10^3\,L_2^r$ &$10^3\,L_3^r$ &$10^3\,L_4^r$
     &$10^3\,L_5^r$ &$10^3\,L_6^r$ &$10^3\,L_7^r$&$10^3\,L_8^r$\\ \hline
(A)   &  1.11 & 1.05 &-3.82 &1.87 &1.22 &1.46 &-0.39 & 0.65 \\
(B)   &  1.00 & 1.48 &-3.82 &0.30 &1.23 &0.14 &-0.27 & 0.55 \\ \hline
\et
\caption{\small Two sets of central values of $L_i^r(\mu)$
  with $\mu=0.77$ GeV from NLO fits performed ref.~\cite{Bijnens:2014lea}. }
\lbltab{BE14LECs}
\end{table}
%%%%%
The values of the low-energy couplings (LEC's) $L_i^r$, $i=1\cdots8 $ are
needed in order to evaluate numerically the chiral amplitudes. A
recent update of the values of the couplings $L_i^r$ has been presented
in ref.~\cite{Bijnens:2014lea} based on global fits involving a number of low
energy observables. We reproduce in table~\Table{BE14LECs} two sets of
values which correspond to NLO expansions (which seem appropriate here
since we are using NLO formulae). The set labelled (A) in
table~\Table{BE14LECs} corresponds to an unconstrained fit and it leads
to rather large values of the  couplings $L_4$, $L_6$
and $L_2-2L_1$ which are suppressed  in the large $N_c$ limit~\cite{gl85}. 
The set (B) in the table corresponds to a fit which is constrained to
enforce compatibility with the results from lattice QCD simulations on
$L_4^r$ and $L_6^r$. We will consider it to be more plausible, since
the strong deviations from the large $N_c$ limit are possibly an artifact of
attempting to reproduce certain observables which are sensitive to
NNLO rescattering effects (like the $I=J=0$ $\pi\pi$ scattering
length) using NLO formulae.
Fig.~\fig{AmplitChir} illustrates the sensitivity of the $I=1$
amplitudes considered here to the  $1/N_c$ suppressed couplings. 
The shape of the $\eta\pi\to\eta\pi$ amplitude is quite different if
one uses the set (A) or the set (B). This is also reflected in the
values of the $J=0$ threshold parameters. Defining the scattering length $a_0$
and the scattering range $b_0$ as in ref.~\cite{Bernard:1991xb}, 
\be
\frac{2}{\sqrt{s}}\,T^{11}(s)= a_0 + b_0\, p^2 +\cdots
\en
with $\sqrt{s}=\sqrt{\mpid+p^2}+\sqrt{\metad+p^2}$, one finds
\be\ba{lll}
m_\pi\,a_0 = 6.7\cdot10^{-3},\quad & m_\pi\,b_0 = -15.0\cdot10^{-3}\quad
 & (\hbox{Large } L_4,\ L_6) \\
 m_\pi\,a_0 = 16.2\cdot10^{-3},\quad & m_\pi\,b_0 = 10.6\cdot10^{-3}\quad
 & (\hbox{Small } L_4,\ L_6) \ .
\ea\en
The two sets of couplings thus lead to rather different values of the
scattering length $a_0$ while the values of the scattering range $b_0$ differ
in their sign. At leading chiral order, one has $m_\pi\,a_0 =
6.2\cdot10^{-3}$, $b_0=0$. At NLO, a low-energy theorem (LET) for $a_0$ was
derived in ref.~\cite{Kubis:2009sb}, in the form of  a linear relation 
\be\lbl{bastianLET}
\left.a_{0}\right\vert_{NLO} = \lambda\,
\left.a_{0,\pi\pi}^2\right\vert_{NLO} +\mu
\en
where $a_{0,\pi\pi}^2$ is the $\pi\pi$ scattering length with $J=0$, $I=2$ and
$\lambda$, $\mu$ are simple functions of the masses $m_\pi$, $m_K$, $m_\eta$
and the decay constants $F_\pi$, $F_K$.  The most precise determinations of
the $S$-wave $\pi\pi$ scattering lengths are based on Roy equations solutions.
Using the values quoted in two recent analysis of these
equations~\cite{Colangelo:2001df,GarciaMartin:2011cn} in the LET
relation~\rf{bastianLET}, one obtains 
\be\lbl{a0LET}
\ba{lll}
a_{0,\pi\pi}^2=-0.0444\pm0.0010 & (\hbox{ref.~\cite{Colangelo:2001df}}) & 
\longrightarrow a_0=(-0.22\pm6.26)\cdot10^{-3} \\
a_{0,\pi\pi}^2=-0.042\pm0.0040 & (\hbox{ref.~\cite{GarciaMartin:2011cn}}) & 
\longrightarrow a_0=(14.8\pm25.0)\cdot10^{-3}\ .
\ea\en
This illustrates that the LET is practically useful only if $a_{0,\pi\pi}^2$
is known to a very high precision. The result of
ref.~\cite{Colangelo:2001df} is associated with a rather small error of
$2.5\%$. However, the result derived from the
Roy equations concerns the physical value of the scattering length rather than
the NLO value which enters into the LET. An additional error should therefore
be introduced in eq.~\rf{a0LET} in order to account for the difference 
$a_{0,\pi\pi}^2-\left.a_{0,\pi\pi}^2\right\vert_{NLO}$, which could
easily be as large than $5\%$. This observation then limits the
effectiveness of the LET for determining $a_0$.

The $\kzerob\kplus\to \kzerob\kplus$ partial-wave amplitude vanishes
at $s=0$ at leading chiral order~\rf{T0ijop2}. This zero, however, is
accidental since it is not associated with a soft pion
theorem. Fig.~\fig{AmplitChir} shows that, indeed, the NLO corrections
are substantial. The corrections corresponding to the $L_i$ set (B),
with small $1/N_c$ violations, have a more reasonable size than those from
set (A). The amplitude
$\eta\pi^+\to \kzerob\kplus$ has a zero at $s=4\mkd/3$ at $O(p^2)$
which corresponds to a soft pion Adler zero. Fig.~\fig{AmplitChir}
shows that the NLO corrections are rather small in this case and that
there is little difference between the couplings of set (A) and set
(B).
%%%%%
%%%%%

\section{Form factors from dispersive integral equations}
We follow here a general approach to the construction of form factors which
implements unitarity relations and chiral constraints and, additionally,
impose the absence of zeros and consistency with the QCD asymptotic
behaviour. We will briefly review this method below, which was applied
previously to the scalar $\pi\pi$ and $\pi K$ form
factors~\cite{Donoghue:1990xh,Jamin:2001zq}, and allows one to relate the form
factors and the corresponding $S$-wave scattering amplitudes via a set of
integral equations. The $I=1$ scalar form factors $F_S^{\eta\pi}$,
$F_S^{K\Kbar}$ which we will discuss here were considered previously in
ref.~\cite{Guo:2012yt}. The  approach followed in ref.~\cite{Guo:2012yt}
differs from ours in that the constraints on the zeros and the asymptotic
behaviour were not imposed. 
 
\subsection{Phase dispersive representation}
The crucial property of two-meson form factors is that they can
be defined as analytic functions in the complex energy plane, with a
cut lying on the positive real axis in the range
$s > (m_P+m_Q)^2$~\cite{Barton:1965}. In the asymptotic region,
$|s|\to\infty$, the general arguments concerning exclusive processes
in QCD~\cite{Lepage:1980fj} predict that a two-meson scalar form
factor $F_S$ should obey a power law behaviour,
\be\lbl{FSasy}
\left.F_S(s)\right\vert_{s\to\infty} \sim 1/s
\en
up to logarithms. Making the assumption that the form factor $F_S$ has no
zeros in the complex plane, one can derive a minimal phase dispersive
representation (e.g.~\cite{Gourdin:1974iq}),
\be\lbl{FSphaserep}
F_S(s)=F_S(0)\,\exp\left[\frac{s}{\pi}\int_{s_0}^\infty 
\frac{\phi_S(s')}{s'\,(s'-s)}\,\,ds'
\right]\ ,
\en 
where the phase is defined from $F(s+i\epsilon)=|F_s(s)|\exp(i\phi_S(s))$.
The QCD asymptotic behaviour~\rf{FSasy} is reproduced from eq.~\rf{FSphaserep}
provided that the phase has the asymptotic limit:
\be
\lim_{s'\to +\infty}\phi_S(s')=\pi\ .
\en
The scalar radius, finally, is given by a simple integral as a
function of $\phi_S$,
\be\lbl{r2phaseint}
\braque{r^2}_S= \frac{6}{\pi}\int_{s_0}^\infty
\frac{\phi_S(s')}{(s')^2}\,\,ds'\ .
\en
If $n$ complex zeros were present, then the right-hand side of
eq.~\rf{FSphaserep} would have to be multiplied by a polynomial of degree $n$ 
and the asymptotic phase would have to be $(n+1)\pi$. The minimality assumption
is equivalent to stating that the increase of the phase in the energy region
$\sqrt{s} >2$ GeV should be less than $\pi$. This is plausible since
no sharp resonances are present in this region.

%% Plausibility argument for absence of zeros ?
%%The equations above are valid when the form factor has no zeros. If,
%%for instance, the form factor has a pair of complex conjugate zeros
%%$z_0$, $z^*_0$, the representation~\rf{FSphaserep} has to be multiplied
%%by $(s-z_0)(s-z_0^*)$ and the asymptotic value of $\phi_S$ would be $3\pi$
%%instead of $\pi$. 
\subsection[Determination of the form factors from the $T$-matrix]{\boldmath Determination of the form factors from the $T$-matrix} 
As emphasised in ref.~\cite{Yndurain:2005cm}, these phase relations are of
particular interest for those form factors which involve at least one pion,
$F_S^{\pi P}$ with $P=\pi$, $K$ or $\eta$ which interests us here. This is
simply because the scattering amplitudes $\pi P\to \pi P$ are \emph{elastic}
in a finite low energy region.  In this region, the form factor phase
$\phi_S^{\pi P}$ is constrained from Watson's theorem to be exactly equal to
the elastic scattering phase shift. The energy region in which inelasticity
can be neglected to a good approximation extends up to the $K\Kbar$ threshold
for $\pi\pi$ and we expect the same property to hold also\footnote{The
  inelastic mode $\eta\pi\to 3\pi$ is allowed already at threshold but the
  $S$-wave projection vanishes by parity conservation (since $J^P=0^-$ for the
  $3\pi$ state). The modes $\eta\pi\to 5\pi$, $\eta\pi\to\eta3\pi$ are
  strongly suppressed by phase space below one GeV.} for $\pi\eta$.  The
asymptotic value of the form factor phase is also known and one may estimate
that $\phi_S^{\pi P}$ should be smoothly approaching its asymptotic value when
$\sqrt{s}\gapprox 2$ GeV. There only remains to determine $\phi_S^{\pi P}$ in
the intermediate energy region that is, in the case of $\eta\pi$, in the
region $1\le \sqrt{s}\lapprox 2$ GeV. In this region, we further expect that
the fastest energy variation should take place close to 1 GeV, associated with
the sharp onset of inelasticity triggered by the presence of the $a_0(980)$
resonance which is known to couple strongly to the $K\Kbar$
channel~\cite{Astier:1967zz}.  This suggests to consider a framework which
takes into account only the dominant inelastic channel and ignores all the
other ones. In this case, the two form factors $F_S^{\eta\pi}$, $F_S^{K\Kbar}$
obey a closed set of Muskhelishvili-Omn\`es coupled integral equations,
\be\lbl{FFintequations}
\begin{pmatrix}
F_S^{\eta\pi}(s)\\
F_S^{K\Kbar}(s)\\
\end{pmatrix}=
\dfrac{1}{\pi}\int_{(m_\eta+\mpi)^2}^\infty \dfrac{ds'}{s'-s}
\begin{pmatrix}
T^{11}(s) & T^{12}(s) \\
T^{12}(s) & T^{22}(s) \\
\end{pmatrix}^*
\begin{pmatrix}
\sigma_1(s')\,F_S^{\eta\pi}(s')\\
\sigma_2(s')\,F_S^{K\Kbar}(s')\theta(s'-4\mkd)\\
\end{pmatrix}\ .
\en
These equations encode the property of analyticity of the form factors, the
asymptotic behaviour (which allows for an unsubtracted dispersive
representation) and two-channel unitarity. One can express the two-channel
$S$-matrix in terms of two phase shifts and one inelasticity parameter in the
usual way,
\be
\bm{S}=\begin{pmatrix}
\eta\,\hbox{e}^{2i\delta_{11}} & 
i\sqrt{1-\eta^2}\, \hbox{e}^{i(\delta_{11}+\delta_{22})} \\
i\sqrt{1-\eta^2}\, \hbox{e}^{i(\delta_{11}+\delta_{22})} & 
\eta\,\hbox{e}^{2i\delta_{22}} \\ 
\end{pmatrix},\quad
0\le \eta \le 1\ .
\en
We assume the following asymptotic conditions on the $S$-matrix parameters
\be\lbl{asycond} 
\lim_{s\to\infty} \eta(s)=1,\quad \lim_{s\to\infty}
\delta_{11}(s)+\delta_{22}(s)=2\pi~, \en
which ensure that the so called Noether index~\cite{FritzNoether}
(see also~\cite{Muskhelishvili}) associated with the set of singular integral
equations~\rf{FFintequations} is equal to two. This, in general, implies that
a unique solution is obtained once two arbitrary conditions are specified, for
instance the values at $s=0$: $F_S^{\eta\pi}(0)$, $F_S^{K\Kbar}(0)$, and that
the solution form factors behave asymptotically as
$1/s$~\cite{Muskhelishvili}.

In summary, solving the set of eqs.~\rf{FFintequations} for the form factors
$F_S^{\eta\pi}$, $F_S^{K\Kbar}$, one obtains a phase 
$\phi_S^{\eta\pi}$ which correctly matches with both the low and high energy limits
expectations and provides an interpolating model in the intermediate energy
region. The phase $\phi_S^{K\Kbar}$ is also provided. In this case, however,
there is no constraint from Watson's theorem at low energy. One expects that
the form factor $F_S^{K\Kbar}$ will be more sensitive than $F_S^{\eta\pi} $ to
the influence of the neglected inelastic channels.

More generally, one can use the system of equations~\rf{FFintequations} to
define the Omn\`es matrix $\Omega^{ij}(s)$ which generalises the usual Omn\`es
function~\cite{Omnes:1958hv}. Such a
generalisation was first discussed in the case of $\pi\pi-K\Kbar$ scattering
in refs.~\cite{Babelon:1976kv,Babelon:1976ww}. The first column of the
Omn\`es matrix is obtained by solving the system with the boundary conditions
$\Omega^{11}(0)=1$, $\Omega^{21}(0)=0$ and the second column by solving with
the conditions $\Omega^{12}(0)=0$, $\Omega^{22}(0)=1$ (see in ref.~\cite{Moussallam:1999aq} an appropriate numerical method for solving the linear system). The Omn\`es matrix allows
one to treat the final-state interaction problem taking into account inelastic
rescattering. For instance, one can express the $I=1$ scalar  form factors
in terms of the $\bm{\Omega}$ matrix,
\be
\begin{pmatrix}
F_S^{\eta\pi}(s)\\
F_S^{K\Kbar}(s)\\
\end{pmatrix}=
\begin{pmatrix}
\Omega^{11}(s) & \Omega^{12}(s)\\
\Omega^{21}(s) & \Omega^{22}(s)\\
\end{pmatrix}
\begin{pmatrix}\,
F_S^{\eta\pi}(0)\\
F_S^{K\Kbar}(0)\\
\end{pmatrix}~.
\en

\section[Two-channel unitary $T$-matrix parametrisation with
chiral matching]{\boldmath Two-channel unitary $T$-matrix parametrisation with
chiral matching} \lblsec{Tmatrixparam}
We seek a parametrisation of the $J=0$ $T$-matrix which: a) should
satisfy exact elastic unitarity below the $K\Kbar$ threshold and exact
two-channel unitarity above, b) should correctly match with ChPT for small
values of $s$  , i.e. 
\be\lbl{chiralmatch} 
T^{ij}(s) - (T^{ij}_{(2)}(s)+ T^{ij}_{(4)}(s)) =O(p^6)\ .  
\en 
and c) should be reasonably simple and
flexible and be able to describe scattering in the low to medium energy region
up to, say $\sqrt{s}\simeq 2$ GeV.  We choose a representation somewhat
similar to that proposed in ref.~\cite{Jamin:2000wn} to describe $J=0$ $\pi K$
scattering, belonging to the family of ``unitary chiral'' approaches. Such
approaches were proposed, in the context of ChPT, firstly in
refs.~\cite{Dobado:1989qm,Dobado:1992ha} and multichannel extensions were
discussed in refs.~\cite{Oller:1997ti,Oller:1997ng} (we refer to the
review~\cite{Oller:2000ma} for a survey and a complete list of references).
There are, however, some drawbacks to these methods. Poles can occur on
physical sheets and, furthermore, the structure of the left-hand cuts is not
quite correct. In particular, the left-hand cut of the chiral $K\Kbar\to
K\Kbar$ amplitude $T^{22}_{(4)}(s)$, which extends up to $s=4(\mkd-\mpid)$ is
propagated to the amplitude $T^{11}$, via the unitarisation method, which
actually spoils the unitarity of $T^{11}$ in the elastic region. While the
resulting unitarity violation is numerically
small~\cite{Guerrero:1998ei,GomezNicola:2001as}, we will prefer here to
maintain exact unitarity at the price of relaxing the matching condition for
the component $T^{22}$.

\begin{figure}
\centering
\includegraphics[width=0.50\linewidth]{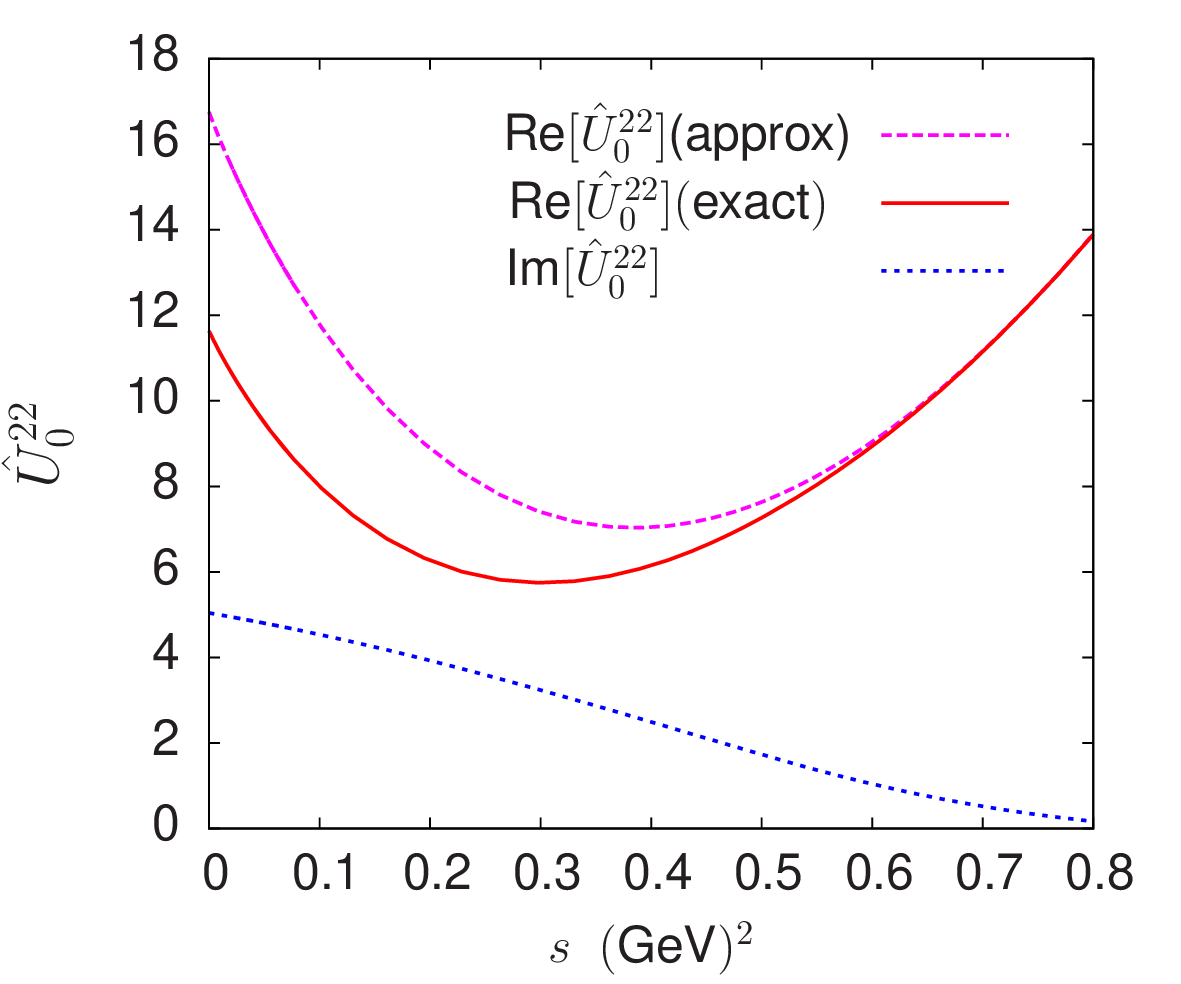}
\caption{\small Comparison of the real part of function $\hat{U}_0^{22}$
with the approximation used in the unitary representation~\rf{Tmodel}. Also
shown is the imaginary part of $\hat{U}_0^{22}$.}
\label{fig:U0hat}
\end{figure}

We start from a  $K$-matrix type representation for the two-channel
$T$-matrix 
\be\lbl{Tmodel} 
\bm{T}(s)=(1- \bm{K}(s)\bm{\Phi}(s))^{-1}
\bm{K}(s)\ .  
\en 
This form is compatible with the symmetry of the
$T$-matrix ($^t\bm{T}=\bm{T}$) provided both $\bm{K}$ and $\bm{\Phi}$
are symmetric matrices. The matrix $\bm{\Phi}(s)$ must also satisfy
\be \lbl{ImPhi}
\im[\bm{\Phi}(s)]=
  \begin{pmatrix}
\theta(s-(m_\eta+m_\pi)^2) \sigma_1(s) & 0 \\ 
0 & \theta(s-4\mkd) \sigma_2(s)\\
\end{pmatrix}
\en 
which ensures that the $T$-matrix obeys the unitarity condition, provided that
the matrix $\bm{K}(s)$ remains real in the range $(m_\eta+m_\pi)^2\le s
  <\infty$.  We take a representation of $\bm{\Phi}(s)$, satisfying
eq.~\rf{ImPhi}, which is diagonal and contains four phenomenological
parameters
\be \bm{\Phi}(s)=
\begin{pmatrix}
\alpha_1+\beta_1 s + 16\pi \bar{J}_{\eta\pi}(s) & 0 \\ 
0 & \alpha_2+\beta_2 s + 16\pi \bar{J}_{K\Kbar}(s)\\
\end{pmatrix}\ .
\en
The parameters $\alpha_i$, $\beta_i$ are assumed to be $O(1)$ in the
chiral counting.  The  $K$-matrix is written in terms of
components with a definite chiral order, \be\lbl{Kmatrix} \bm{K}(s)=
\bm{K}_{(2)}(s)+{\bm{K}}_{(4)}(s)+ \bm{K}_{(6)}(s) \en where, as
before, the subscript denotes the chiral order. In order to satisfy
the matching condition~\rf{chiralmatch} one must have, 
\be
\bm{K}_{(2)}(s)= \bm{T}_{(2)}(s),\quad \bm{T}_{(4)}(s)=
   {\bm{K}}_{(4)}(s) +\bm{T}_{(2)}(s) \bm{\Phi}_{(0)}(s)\bm{T}_{(2)}(s)\ .
\en
One can then express ${\bm{K}}_{(4)}$ in terms of the polynomial and
left-cut functions defined from
eqs.~\rf{T4leftright}~\rf{U0hatdef}~\rf{U0decomp} (see also
appendix~\sect{Decompp4}) 
\be
{\bm{K}}_{(4)}(s)=\frac{1}{16\pi}\,\left( \bm{P}_0(s) +
\hat{\bm{U}}_0(s) \right) -\bm{T}_{(2)}(s)
\begin{pmatrix}
\alpha_1 &0 \\ 0 & \alpha_2\\
\end{pmatrix}
\bm{T}_{(2)}(s)\ .  
\en 
As explained above, we must use an approximation to the function
$\hat{U}_0^{22}$ which has no cut on the real axis in the range $s \ge
(m_\eta+m_\pi)^2 $.  This may be done by removing the parts which are
proportional $\bar{J}_{\pi\pi}(t)$ and $\bar{J}_{\eta\pi}(t)$ (see
eqs.~\rf{V0andV1}) from the two functions $V_0(t)$ and $V_1(t)$, which appear
in the angular integral which gives $\hat{U}_0^{22}$ (see
eq.~\rf{U0hatdef}). Figure~\fig{U0hat} compares this approximation of
$\hat{U}_0^{22}$ to the exact function.

%%%%%
%%%%%

\begin{figure}
\centering
\includegraphics[width=0.499\linewidth]{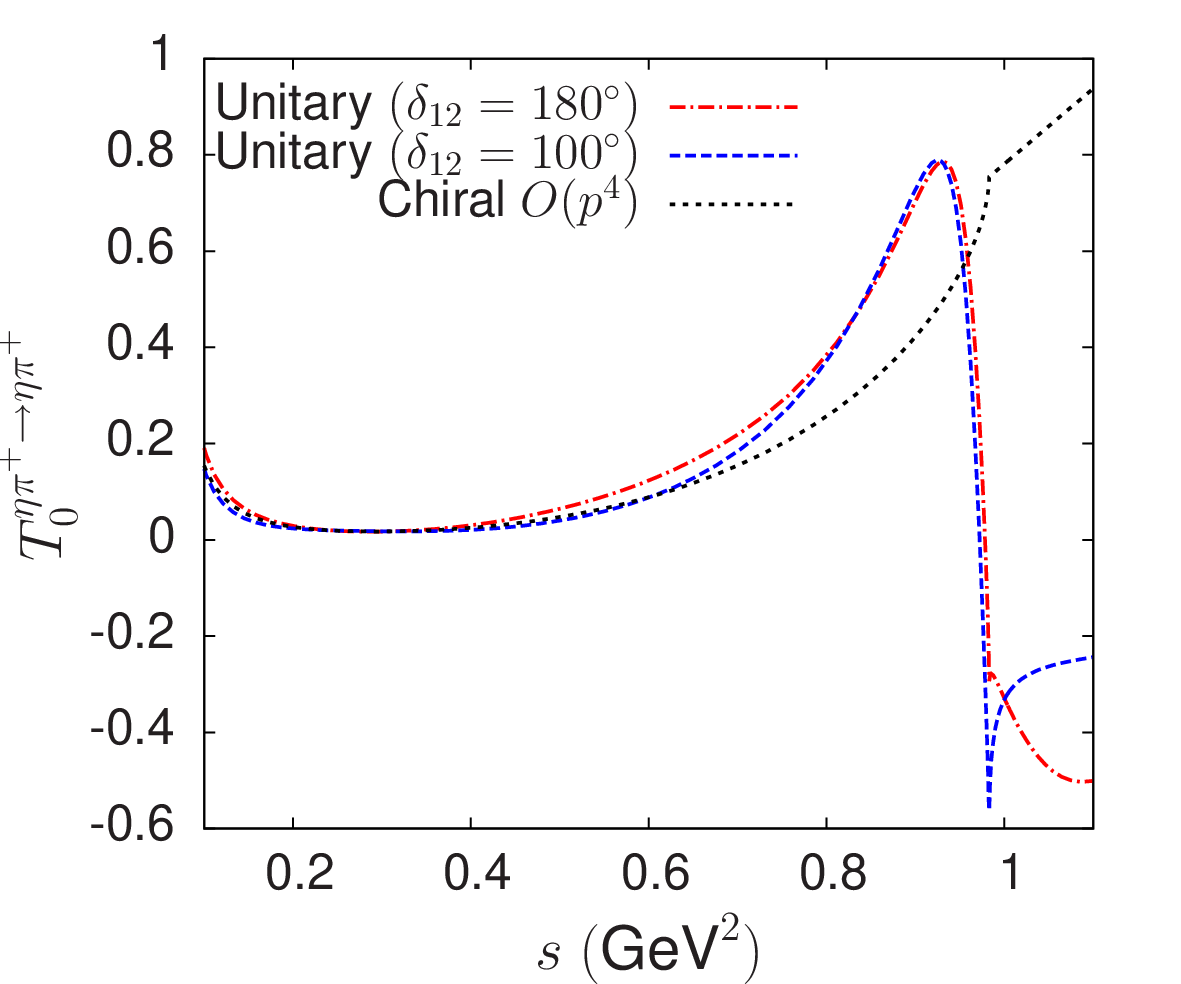}\includegraphics[width=0.499\linewidth]{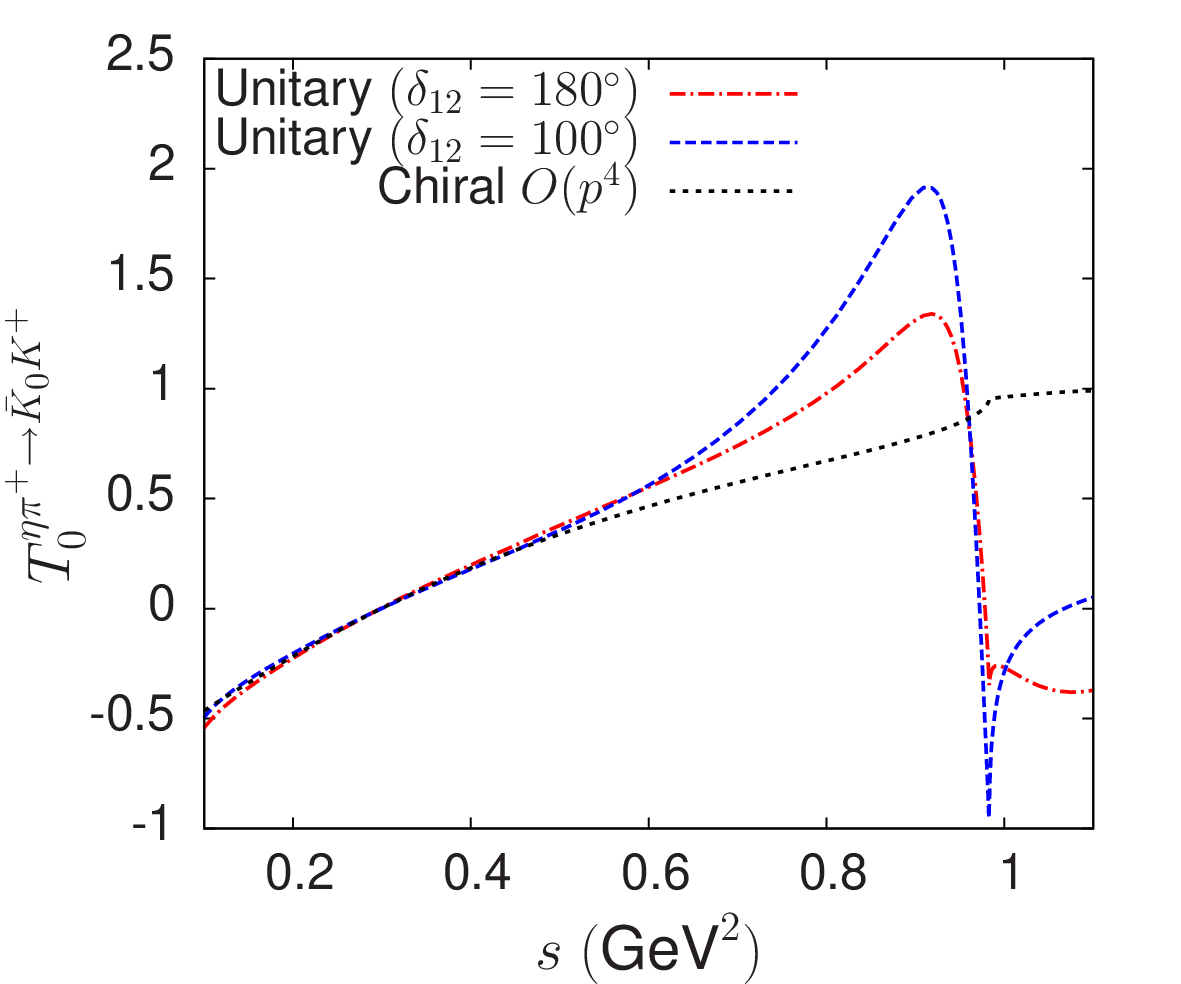}\\ \includegraphics[width=0.5\linewidth]{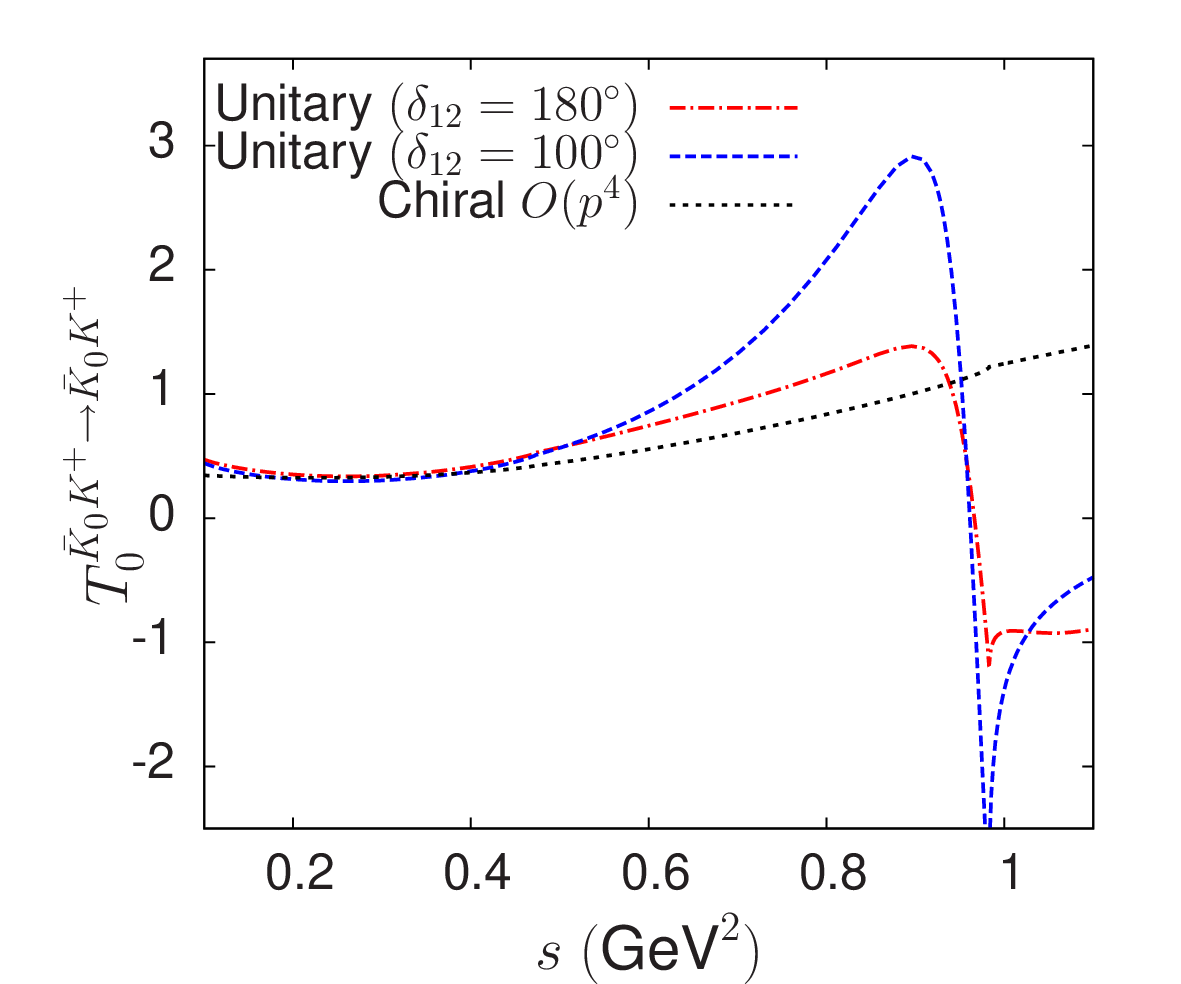}
\caption{\small Comparison of the real parts of unitary partial-wave
  amplitudes $T^{ij}$ given from eq.~\rf{Tmodel} and the corresponding
  chiral amplitudes at NLO.}
\label{fig:AmplitUnit}
\end{figure}

Finally, $\bm{K}_{(6)}(s)$ is taken to be a pole term with the
$O(p^4)$ part removed, 
\be {K}^{ij}_{(6)}(s)= \frac{g_i g_j}{16\pi}
\left( \frac{1}{m_8^2-s} - \frac{1}{m_8^2} \right) \en
We model the couplings $g_1$, $g_2$ such that they behave as $O(p^2)$, based
on a scalar resonance chiral Lagrangian analogous to the one introduced in
ref.~\cite{Ecker:1988te}
\begin{align}
g_1= & \frac{\sqrt6}{3\fpid}( c'_d\,(s-\metad-\mpid)+2c'_m\,\mpid
)~,\nonumber\\
g_2= & \frac{1}{\fpid}( c'_d\,(s-2\mkd)+2c'_m\,\mkd)\ .
\end{align}
We will discuss in sec.~\sect{phenodeterm} how the phenomenological parameters
may be determined from experimental information on the properties of the
$a_0(980)$, $a_0(1450)$ resonances as well as chiral constraints on the
amplitudes and on the $I=1$ scalar form factor. Figure~\fig{AmplitUnit}
illustrates how the unitary amplitudes parametrised as described above
correctly match with the NLO chiral amplitudes at low energy.

%%%%%updated May 8
%%%%%
\section[Phenomenological determination of the phase shifts and inelasticity
  and the $I=1$ scalar form factor]{\boldmath Phenomenological determination of the phase shifts and inelasticity
  and the $I=1$ scalar form factor}\lblsec{phenodeterm}
\subsection[Experimental information on $\pi\eta \to \pi\eta$ and $\pi\eta\to
  K\Kbar$ scattering]{\boldmath Experimental information on $\pi\eta \to \pi\eta$ and $\pi\eta\to
  K\Kbar$ scattering}
Let us first consider the $\pi\eta\to \pi\eta$ amplitude below the $K\Kbar$
threshold. In this region, $\eta\pi$ scattering should be approximately
elastic.
The $\pi\eta$ scattering phase shift below 1 GeV should be controlled by the
values of the threshold parameters $a_0$, $b_0$ on the one hand and the
properties of the $a_0(980)$ resonance on the other. We will consider that the
values of $a_0$, $b_0$ corresponding to the set of $L_i's$ with small $L_4$,
$L_6$ (set (B), see table~\Table{BE14LECs}) are the most plausible. In this
case, $a_0$ and $b_0$ are both positive and one expects that the phase shift
will be positive in the whole elastic region. A different possibility was
investigated in ref.~\cite{Achasov:2010kk}.

The $a_0(980)$ is a well established resonance but its shape is not well
described by a simple Breit-Wigner form because of the vicinity of the
$K\Kbar$ threshold. This partly explains the dispersion in the values of the
mass and width quoted by the PDG~\cite{Agashe:2014kda}: $m_{a_0}=980\pm20$
MeV, $\Gamma_{a_0}=[50-100]$ MeV.
A comparison of a number of determinations of the
$T_{11}$ amplitude near the $K\Kbar$ threshold based, in particular, on the
popular Flatt\'e model~\cite{Flatte:1976xu} is performed in
ref.~\cite{Baru:2004xg}. The corresponding $\eta\pi$ phase shifts are plotted
on Fig. 10 of that reference, from which one can deduce that the value of the
phase shift at the $K\Kbar$ threshold lies around 90\degree,
\be
\delta_{11}(2m_K)=(90\pm20)\degree\ .
\en
This is also satisfied in the models of refs.~\cite{Black:1999dx} and
~\cite{Furman:2002cg} which give, respectively,
$\delta_{11}(2m_K)=95\degree$ and $\delta_{11}(2m_K)=77\degree$.  

The $a_0(980)$ resonance corresponds to poles of the amplitude in the
complex plane on the second and on the third Riemann sheets which can
both be near the physical region since the mass is very close to the
$K\Kbar$ threshold.  For definiteness, we will rely here on the recent
determination by the KLOE collaboration~\cite{Ambrosino:2009py}. It is
based on measurements of the $\phi\to \eta\pi\gamma$ decay amplitude
with both high precision and high statistics. Based on the best fit
performed in ref.~\cite{Ambrosino:2009py} (using the theoretical model
from ref.~\cite{Isidori:2006we}) the location of the poles can be
deduced to be
\be\ba{ll}
 \sqrt{s^{II}_{a_0(980)}}= & (994\pm 2 -i\,(25.4\pm 5.0) )\ \hbox{MeV}
 \\[0.25cm] 
 \sqrt{s^{III}_{a_0(980)}}=&  (958\pm 13 -i\,(60.8\pm11.5) )\ \hbox{MeV}\ . \\
\ea\en
In the $[1-2]$ GeV energy region,
a second resonance, the $a_0(1450)$, first reported in
ref.~\cite{Boutemeur:1989qq} was later identified in $\bar{p}p$ decays at rest
(e.g.~\cite{Amsler:1994pz,Abele:1998qd,Bertin:1998sb}, see
also~\cite{Bugg:2008ig} who re-analysed the data). This resonance should
correspond to a pole on the third Riemann sheet. Based on the value of the
mass and width quoted in the PDG, we can estimate
\be
\sqrt{s_{a_0(1450)}^{III}}=(1474\pm19 -i\,(133\pm7) )\ \hbox{MeV}\ .
\en 
A further property of the $a_0(1450)$ is that it has approximately equal decay
widths into $\pi\eta$ and into $K\Kbar$. We will implement this feature by
requiring that the $J=0$ cross sections for $\eta\pi\to \eta\pi$ and
$\eta\pi\to K\Kbar$  should be approximately equal when $\sqrt{s}=1.474$
GeV. In our two-channel framework, these  cross sections have the
following expressions in terms of the phase shifts and the inelasticity
parameter
\be
\sigma(\eta\pi\to \eta\pi)=\frac{\pi}{p_{\eta\pi}^2}
\left \vert \eta\,\hbox{e}^{2i\delta_{11}} -1 \right\vert^2,\quad
\sigma(\eta\pi\to K\Kbar)= \frac{\pi}{p_{\eta\pi}^2}\left(1-\eta^2 \right)
\en
and we expect that $\eta$ should reach a minimum at the mass of the
$a_0(1450)$ resonance. If the minimum is close to zero, the two cross sections
will be approximately equal\footnote{Equality of the two cross sections occurs
either when $\eta=0$ or $\eta=\cos2\delta_{11}$}. In this situation, we
expect a rapid variation of the phase shifts $\delta_{11}$, $\delta_{22}$
(possibly becoming discontinuous if $\eta=0$) at the energy
$\sqrt{s}=m_{a_0(1450)}$. In contrast, the sum of the two phase shifts (which
is also the phase of $S_{12}$) should be a smoothly varying
function. It is convenient to characterise the global behaviour of the
$S$-matrix in the $[1-2]$ GeV region in terms of the value of this
phase sum $\delta_{11}+\delta_{12}$ when $\sqrt{s}=m_{a_0(1450)}$
\be\lbl{defdelta12}
\delta_{12}\equiv  \left. \delta_{11}(\sqrt{s})
+\delta_{22}(\sqrt{s})
\right\vert_{\sqrt{s}=m_{a_0(1450)}}\ .
\en

%%%%% plots updated May 8
\begin{figure}
\centering
\includegraphics[width=0.499\linewidth]{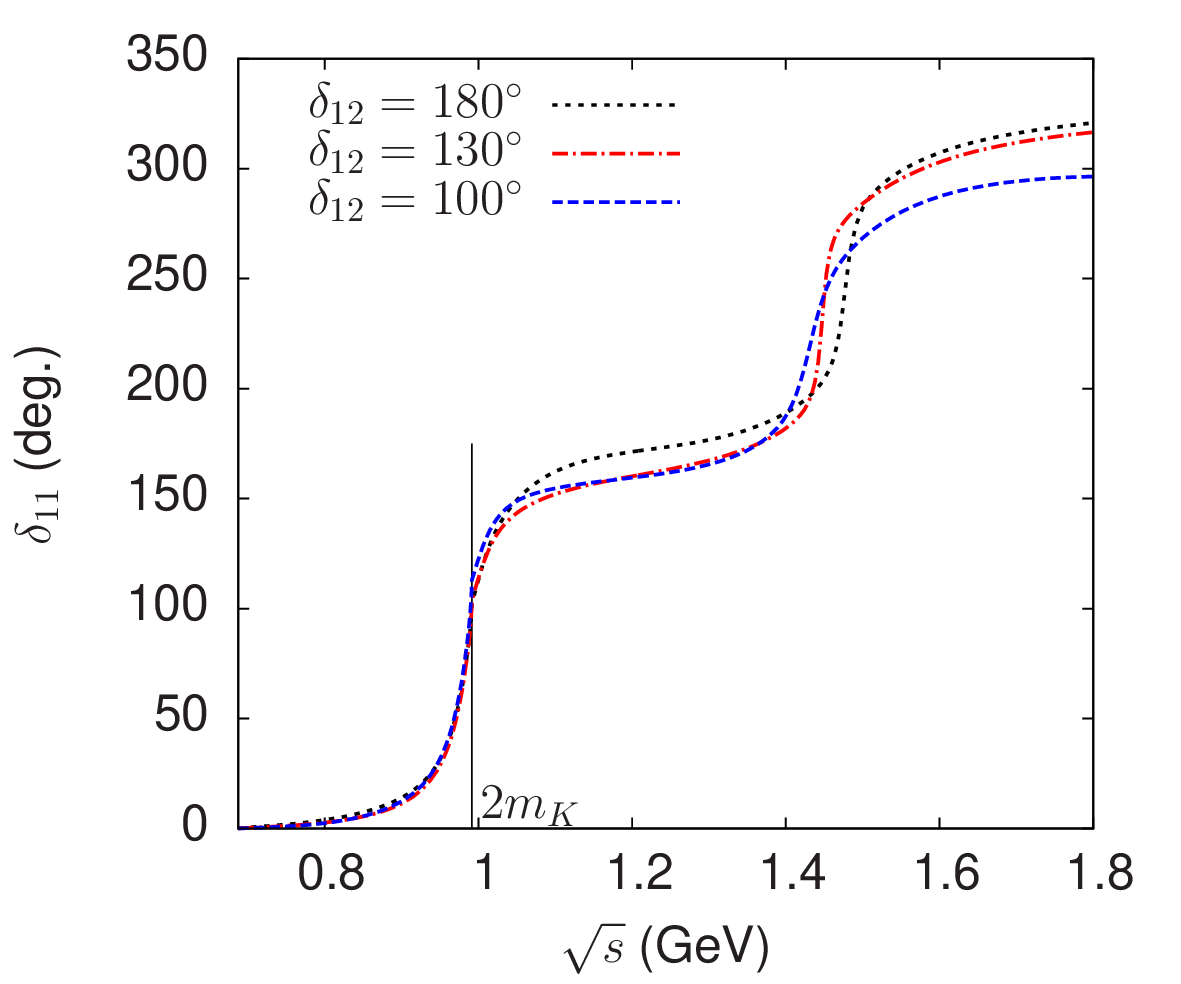}\includegraphics[width=0.499\linewidth]{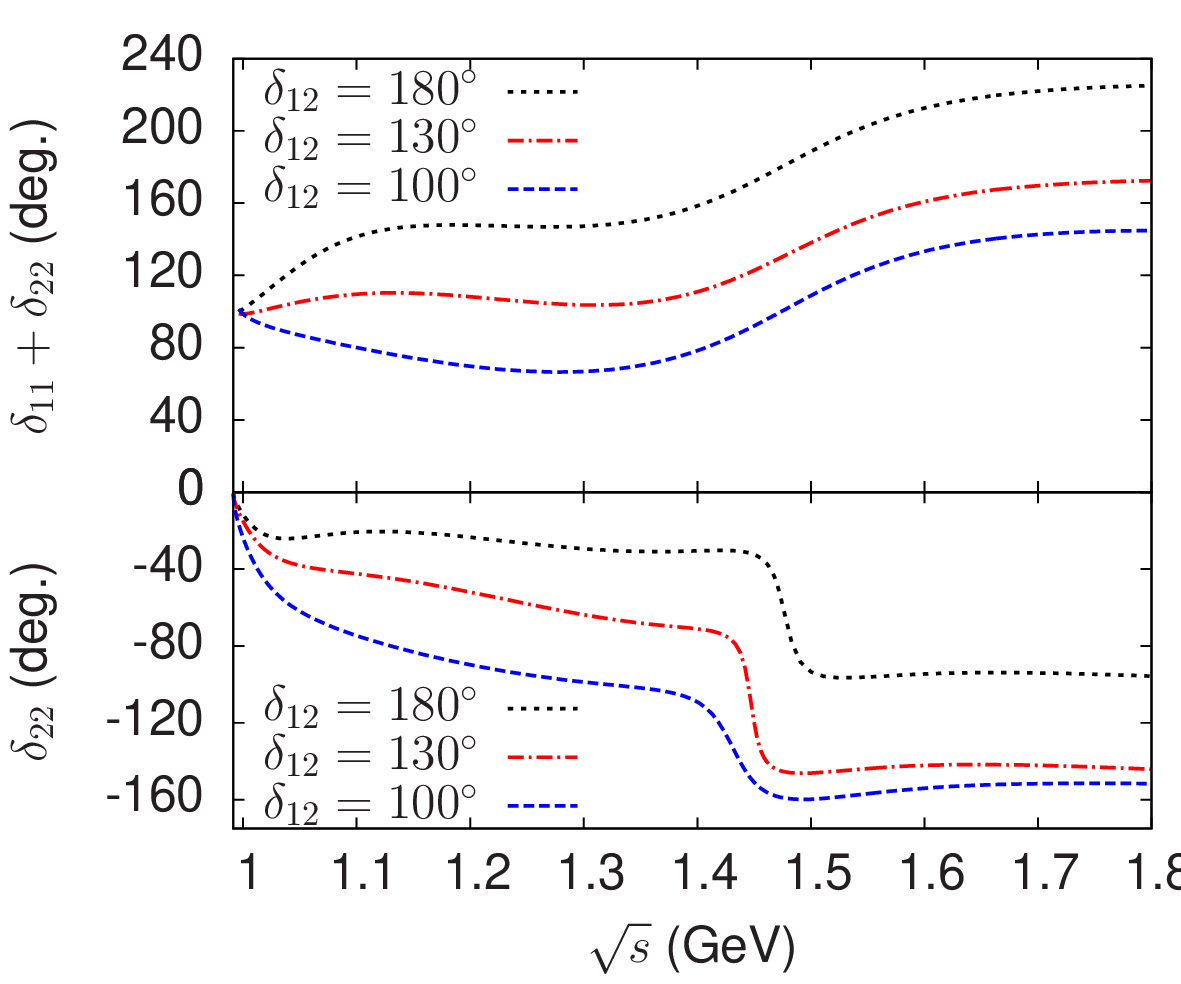}\\ \includegraphics[width=0.5\linewidth]{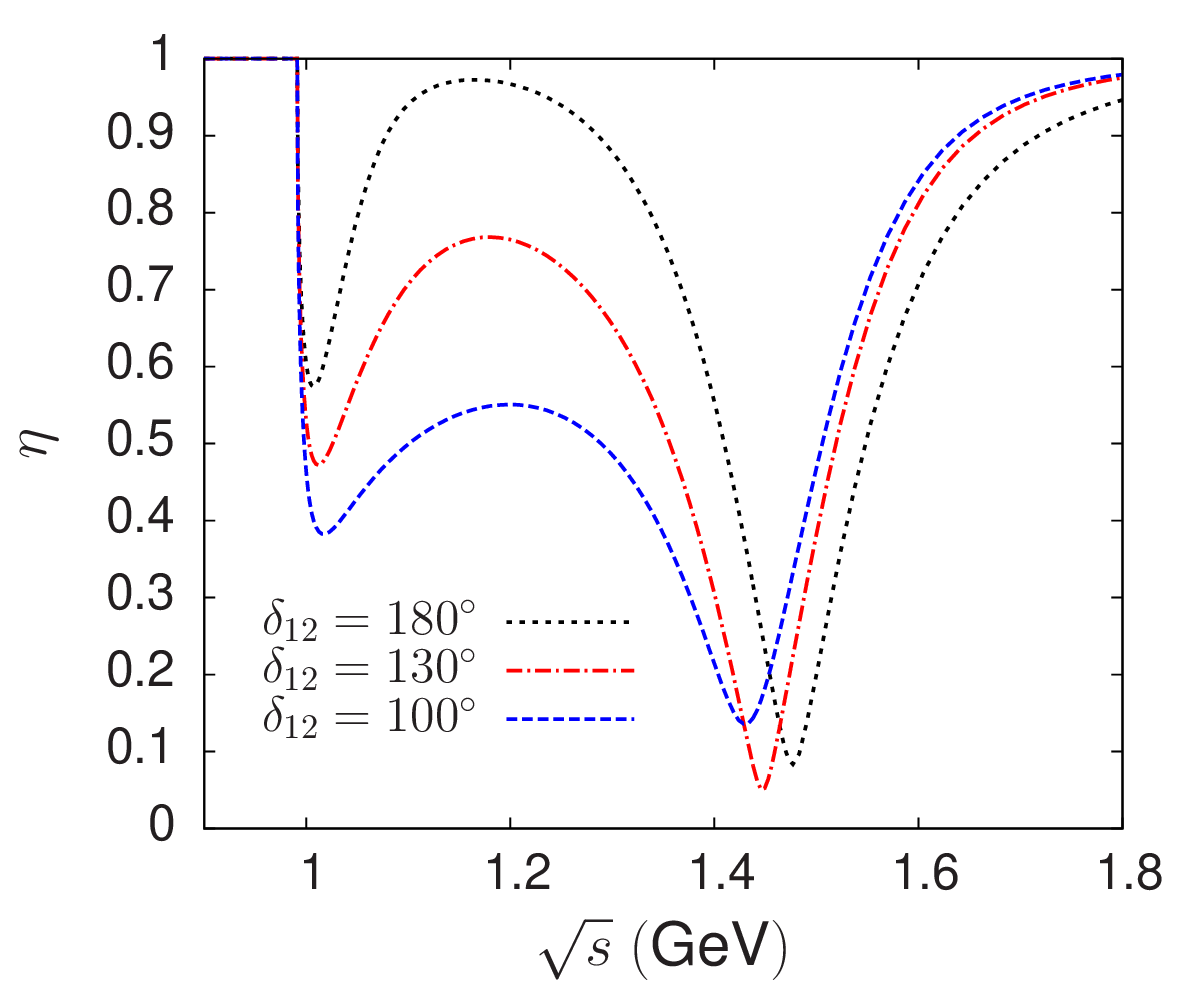}
\caption{\small Phases $\delta_{11}$, $\delta_{22}$, their sum and the
  inelasticity $\eta$ from the $T$-matrix model of sec.~\sect{Tmatrixparam}
corresponding to several imposed values of $\delta_{12}$ (defined in
eq.~\rf{defdelta12}).} 
\label{fig:deltaplot}
\end{figure}
%%%%%
Let us now return to the parametrisation of the $T$-matrix described
in sec.~\sect{Tmatrixparam}. The $T$-matrix elements in this model
have analyticity properties and can be defined away from the physical
region, in the complex energy plane. Using eq.~\rf{Tmodel}, the poles of the
$T$-matrix correspond to the zeros of the determinant
\be
\Delta(s)=\hbox{det}[ 1-\bm{K}(s)\bm{\Phi}(s)]\ .
\en
Recalling that the extension of the loop functions $\bar{J}_{PQ}$ to
the second Riemann sheet are defined as
\be
\bar{J}_{PQ}^{II}(s)= \bar{J}_{PQ}(s) + \frac{i\,\sqrt{\lambda_{PQ}(s)}}{8\pi\,s}
\en
then, the extension of the $T$-matrix elements to the second Riemann sheet is
performed by replacing $\bar{J}_{\eta\pi}(s)$ by $\bar{J}_{\eta\pi}^{II}(s)$
in the matrix $\bm{\Phi}$. Similarly, the extension to the third Riemann sheet
is performed by replacing both $\bar{J}_{\eta\pi}$ and $\bar{J}_{K\Kbar}$ by
$\bar{J}_{\eta\pi}^{II}$ and $\bar{J}_{K\Kbar}^{II}$ in $\bm{\Phi}$.

This $T$-matrix model involves the phenomenological parameters: $\alpha_1$,
$\alpha_2$, $\beta_1$, $\beta_2$, $m_8$, $c'_d$, $c'_m$. For simplicity, we
will keep the ratio $c'_m/c'_d$ fixed and allow only six parameters to
vary. We determine them by imposing six conditions on the $T$-matrix:
\begin{itemize}
\item[a)]As first four conditions, we impose that the real and imaginary
  parts of the poles ${s^{II}_{a_0(980)}}$ and ${s^{III}_{a_0(1450)}}$
  be  reproduced.
\item[b)]As a fifth condition, we impose that the minimum of the
  inelasticity parameter at $\sqrt{s}=m_{a_0(1450)}$ be close to zero
  (in practice, we used $\eta_{min}\approx0.05$, as in
  ref.~\cite{Furman:2002jp}).
\item[c)] As a final condition, we choose a value for the phase 
  $\delta_{12}$ as defined in eq.~\rf{defdelta12}.
\end{itemize}
Within this model, having imposed the first five conditions, the value of
$\delta_{12}$ is found to be bounded from above: $\delta_{12}\lapprox
205\degree$. In addition, consistently with our assumption that most of the
phase variations should take place below 2 GeV, it seems plausible that the
phase sum $\delta_{11}+\delta_{22}$ should not be smaller than its value at
the mass of the $a_0(980)$, i.e. one should have $\delta_{12}\gapprox
90\degree$. 
Fig.~\fig{deltaplot} shows results from this model for the phases
$\delta_{11}$, $\delta_{22}$ and the inelasticity $\eta$ as a function of
energy, corresponding to several different imposed values of $\delta_{12}$.
One observes that the two phases $\delta_{11}$, $\delta_{22}$ undergo a sharp
variation, in opposite directions, close to the mass of the $a_0(1450)$
resonance. The figure illustrates a pattern where $\delta_{11}$ increases
while $\delta_{22}$ decreases. However, a small modification of the
phenomenological parameters which enter into the $T$-matrix model can lead to
a pattern with a reversed behaviour (with $\delta_{11}$ decreasing and
$\delta_{22}$ increasing) which would then be similar to the one obtained in
ref.~\cite{Furman:2002jp}. In contrast, the phase sum,
$\delta_{11}+\delta_{22}$ is completely stable and always increases smoothly
as an effect of the resonance. This ambiguity, which can be viewed as a
$\pm\pi$ ambiguity in the individual definition of $\delta_{11}$ and
$\delta_{22}$ does also not affect observables, in particular, the
determination of the form factors.

\begin{table}[htb]%Updated May 22
\centering
\bt{c|cccccc}\hline\hline
$\delta_{12}$ & $\alpha_1$ & $\alpha_2$ 
& $\beta_1\,(\hbox{GeV}^{-1})$ & $\beta_2\,(\hbox{GeV}^{-1})$ &
$m_8\,(\hbox{GeV})$ & $\lambda$ \\ \hline
$200\degree$ &0.6265&     0.0988&   0.2495&    0.1476&      1.0571 &    0.5704
\\ 
$175\degree$ &0.7427&     0.0781&   0.3085&   -0.0590&      1.0913 &    0.8176
\\ 
$150\degree$ &0.8444&     0.0467&   0.2773&   -0.2085&      1.1258 &    1.1017 
\\  
$125\degree$ &0.8765&     0.0016&   0.2134&   -0.3606&      1.1834 &    1.6856
\\   
$100\degree$ &1.0993&    -0.5055&  -0.0358&   -0.2722&      1.5130 &    5.7024
\\ \hline\hline
\et
\caption{\small Parameters of the $T$-matrix model corresponding to five fixed
  conditions (see text) and several input values of the phase
  $\delta_{12}$. The parameters $c'_m$, $c'_d$ are given in terms of $\lambda$
  by $c'_d= \lambda c_d^0$, $c'_m= \lambda c_d^0/2$ with $c_d^0= 28$ MeV.  }
\lbltab{valparams}
\end{table}

Numerical values for the set of six parameters $\alpha_i$, $\beta_i$, $m_8$,
$c'_d$ corresponding to several input values of $\delta_{12}$ in the range
$90\degree\le \delta_{12}\le 205\degree$ are given in
table~\Table{valparams}. The $T$-matrix is not very sensitive to the value of
the parameter $c'_m$. Very similar results are obtained if one sets $c'_m=0$
or $c'_m=c'_d$. The numerical results shown in the table correspond to taking
$c'_m=c'_d/2$. In this model, the pole of the $K$-matrix corresponds to two
physical resonances. Table~\Table{valparams} shows that the mass parameter of
the pole, $m_8$, varies between 1 and 1.5 GeV, while the value of the parameter
$c'_d$ varies in a rather large range from 16 to 160 MeV, depending on the
input value of the phase $\delta_{12}$.

%%%%%
\begin{figure}
\centering
\includegraphics[width=0.499\linewidth]{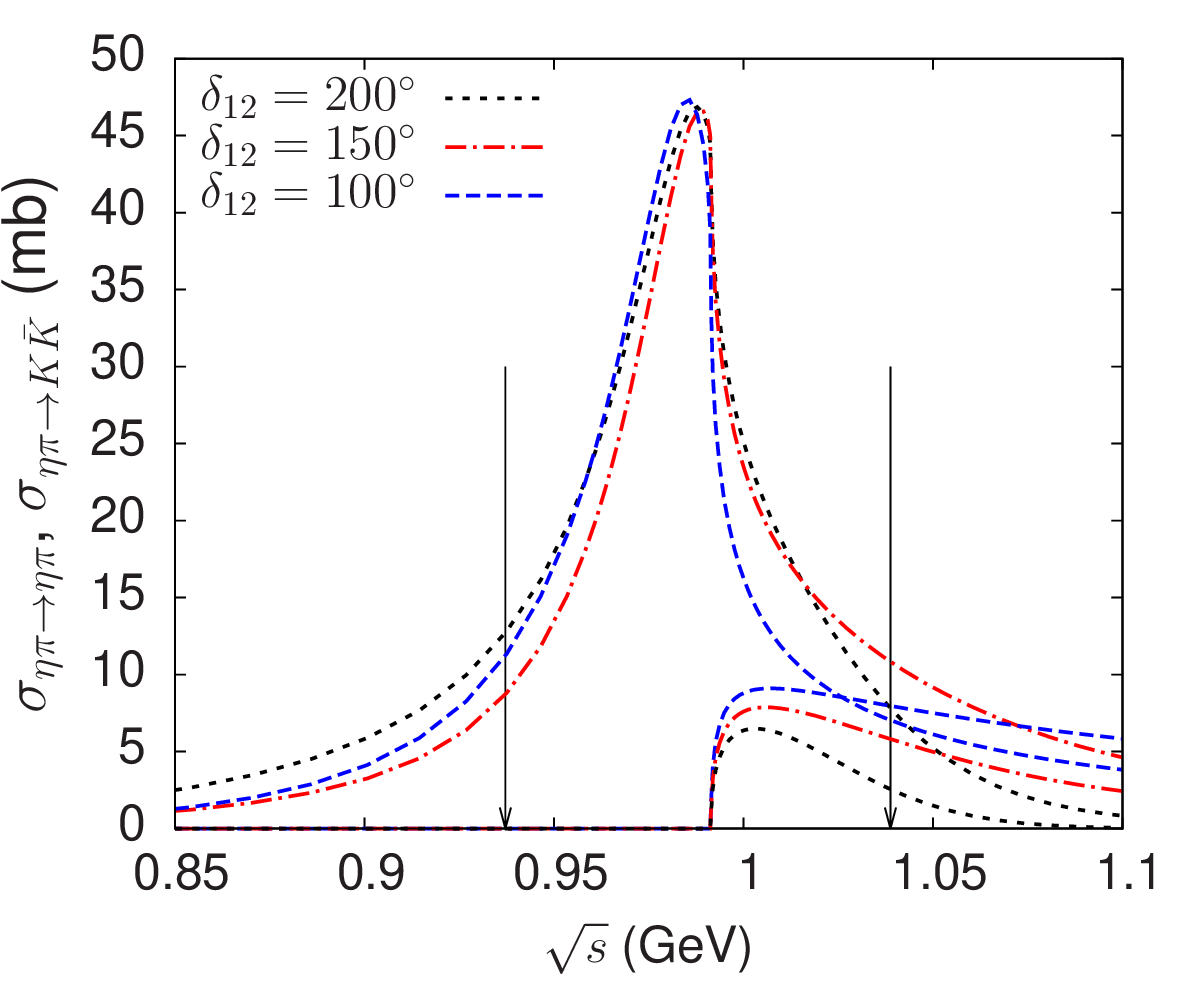}
\caption{\small Cross sections for $\eta\pi\to \eta\pi$ and $\eta\pi\to K\Kbar$
  in the vicinity of the $a_0(980)$ resonance from the $T$-matrix model,
  depending on the input value of $\delta_{12}$. The arrows show the
  integration limits used to define the branching fraction~\rf{bfcrossdef}.}
\label{fig:crosssect}
\end{figure}
%%%%%
The properties of the $a_0(980)$ resonance (apart from the pole position on
the second Riemann sheet which is held fixed) depend on the value of
$\delta_{12}$. Figure~\fig{crosssect} shows the two cross sections
$\sigma_{\eta\pi\to \eta\pi}$, $\sigma_{\eta\pi\to K\Kbar}$ in the vicinity of
the $a_0(980)$ resonance peak. We estimate the branching fraction
$B_{K\Kbar/\eta\pi}= \Gamma_{a_0\to K\Kbar}/\Gamma_{a_0\to \eta\pi}$ in a simple
way in terms of integrals over these cross sections
\be\lbl{bfcrossdef}
B_{K\Kbar/\eta\pi}=\dfrac
{\displaystyle\int_{E^-}^{E^+}   
\sigma_{\eta\pi\to K\Kbar}(E) \,dE}
{\displaystyle\int_{E^-}^{E^+}    
\sigma_{\eta\pi\to \eta\pi}(E) \,dE}
\en
with $E^\pm= m_{a_0}\pm \Gamma_{a_0}$. In this formula, we set $m_{a_0}=988$
MeV, which corresponds to the resonance peak in the cross sections and
$\Gamma_{a_0}=50.8$ MeV corresponding to twice the imaginary part of the pole
position. We collect in table~\Table{a0props} the results for the branching
fraction corresponding to different input values of $\delta_{12}$. The
agreement with the experimental average quoted in the PDG,
$B^{exp}_{K\Kbar/\eta\pi}=0.183\pm0.024$ is qualitatively reasonable, in
particular for the smaller values of $\delta_{12}$. We also indicate in the
table the positions of the $a_0(980)$ pole on the third Riemann sheet (recall that the pole position on the second Riemann sheet is fixed), which
is seen to move away from the real axis as $\delta_{12}$ is decreased.
\begin{table}[h] %updated 5 may
\centering
\bt{c|c|l } \hline\hline
\TT $\delta_{12}$ & $B_{\eta\pi/K\Kbar}$ & $\sqrt{ s^{III}_{a_0}}\,(\hbox{MeV})$ 
\\ \hline
\TT $200\degree$ & 0.095 &  $1022-i\,62$ 
\\
$175\degree$ & 0.127 &  $1020-i\,93$ 
\\
$150\degree$ & 0.148 &  $1009-i\,129$ 
\\
$125\degree$ & 0.170 &  $972 -i\,192$
\\
$100\degree$ & 0.187 &  $749 -i\,376$ 
\\ \hline\hline
\et
\caption{\small Some properties of the $a_0(980)$: values of the
  $\eta\pi/K\Kbar$ branching fraction and position of the pole on the third
  Riemann sheet depending on the input value of the phase $\delta_{12}$.}
\lbltab{a0props}
\end{table}

%%%%%
\subsection[Scalar form factors and the $\eta\pi$ scalar radius]{\boldmath Scalar form factors and the $\eta\pi$ scalar radius}
In order to solve the integral equations~\rf{FFintequations} we must also
define $\delta_{11}(s)$, $\delta_{22}(s)$, $\eta(s)$ for energies above the
mass of the $a_0(1450)$ resonance such that the asymptotic
conditions~\rf{asycond} are satisfied. For this purpose, we define a mapping
$u(s)$ such that $0\le u \le 1$ when $s_1 \le s \le \infty$ and then perform
simple polynomial interpolations of the functions $\delta_{11}$,
$\delta_{22}$, $\eta$ in terms of the variable $u$ (see
appendix~\sect{asympinterp} for more details, in practice we used
$\sqrt{s_1}=1.8$ GeV). For a given value of the phase $\delta_{12}$, the
$T$-matrix is completely specified and one can derive the two scalar form
factors by solving eqs.~\rf{FFintequations}.

%%%%%%% [updated fig. May 5]
\begin{figure}
\centering
\includegraphics[width=0.55\linewidth]{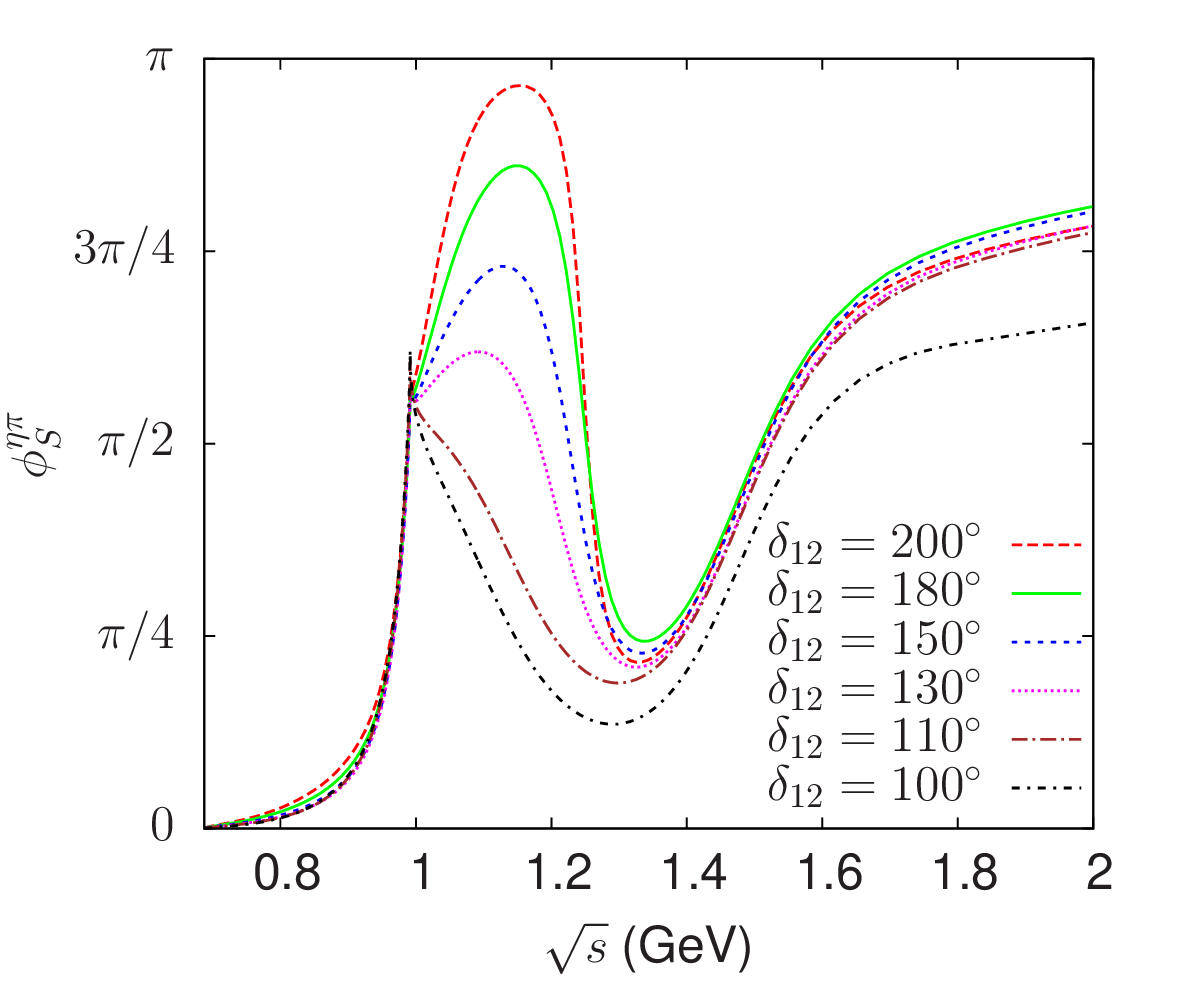}
\caption{\small Phase of the form factor $F_S^{\eta\pi}$ obtained from solving
  the integral equations~\rf{FFintequations} with several input values of the
  phase $\delta_{12}$ (see eq.~\rf{defdelta12}) in the $T$-matrix.}
\label{fig:PhiSetapi}
\end{figure}
%%%%%%%
The form factors turn out to be rather sensitive to the value of
$\delta_{12}$. Fig.~\fig{PhiSetapi} illustrates the numerical results for the
phase of the $\eta\pi$ scalar form factor, $\phi_S^{\eta\pi}$, corresponding
to different input values of $\delta_{12}$. The phase of the form factor
displays a dip located in between the two $a_0$ resonances. This behaviour is
qualitatively similar to the one observed for the scalar form factor phases in
the cases of the $\pi\pi$ or $K\pi$. A detailed discussion can be found in
ref.~\cite{Ananthanarayan:2004xy}.  The phase $\phi_S^{\eta\pi}$ displays a
bump, before the dip, which disappears when the input value of $\delta_{12}$
is smaller than $\simeq130\degree$.  Given the phase integral
representation~\rf{r2phaseint}, we expect the $\eta\pi$ scalar radius to
decrease when $\delta_{12}$ decreases. Numerical values of the scalar radii
for the $\eta\pi$ and the $K\Kbar$ form factors are displayed in
table~\Table{r2numvals} for given values of $\delta_{12}$ in the range
$[100\degree-200\degree]$. In all cases, the dispersive result for
$\braque{r^2}^{\eta\pi}_S$ exceeds the $O(p^4)$ chiral value~\rf{r2Op4numvals}
(the same also holds for the $K\Kbar$ scalar radius). However, one must also
take into account the chiral corrections of order $p^6$ (or higher), the
typical size of which can be as large as $20-30\%$. In the dispersive
evaluation, even if the $T$-matrix elements were known exactly below 2 GeV, an
error would arise from the asymptotic region. This is easily seen from the
phase integral expression~\rf{r2phaseint}. The contribution to the $\eta\pi$
scalar radius from the integration region region $\sqrt{s'}>2$ GeV is
relatively large $\simeq 30\%$ and this could generate an overall uncertainty
for $\braque{r^2}_S^{\eta\pi}$ of the order of $15\%$. The conclusion, then,
is that the chiral result and the dispersive evaluation can be perfectly
compatible provided the phase $\delta_{12}$ lies in the following restricted
range: $90\degree \lapprox \delta_{12} \lapprox 125\degree$.

\begin{table}[h]%updated May 5/May 22
\centering
\bt{ c|ccccc}\hline\hline
$\delta_{12}$ & $200\degree$ & $175\degree$ &  $150\degree$ & 
$125\degree$ & $100\degree$ \\ \hline
\TT $\braque{r^2}^{\eta\pi}_S\,(\hbox{fm}^2)$ & 
0.185 & 0.176 & 0.166 & 0.150 & 0.122
\\ \hline
\TT $\braque{r^2}^{K\Kbar}_S\,(\hbox{fm}^2)$ &
0.253 & 0.248 & 0.245 & 0.233 & 0.209 
\\ \hline\hline
\et
\caption{\small Results for the scalar radii obtained from solving
  eqs.~\rf{FFintequations} for the form factors depending on the input value
  for the phase $\delta_{12}$.}
\lbltab{r2numvals}
\end{table}

\begin{figure}
\centering
\includegraphics[width=0.499\linewidth]{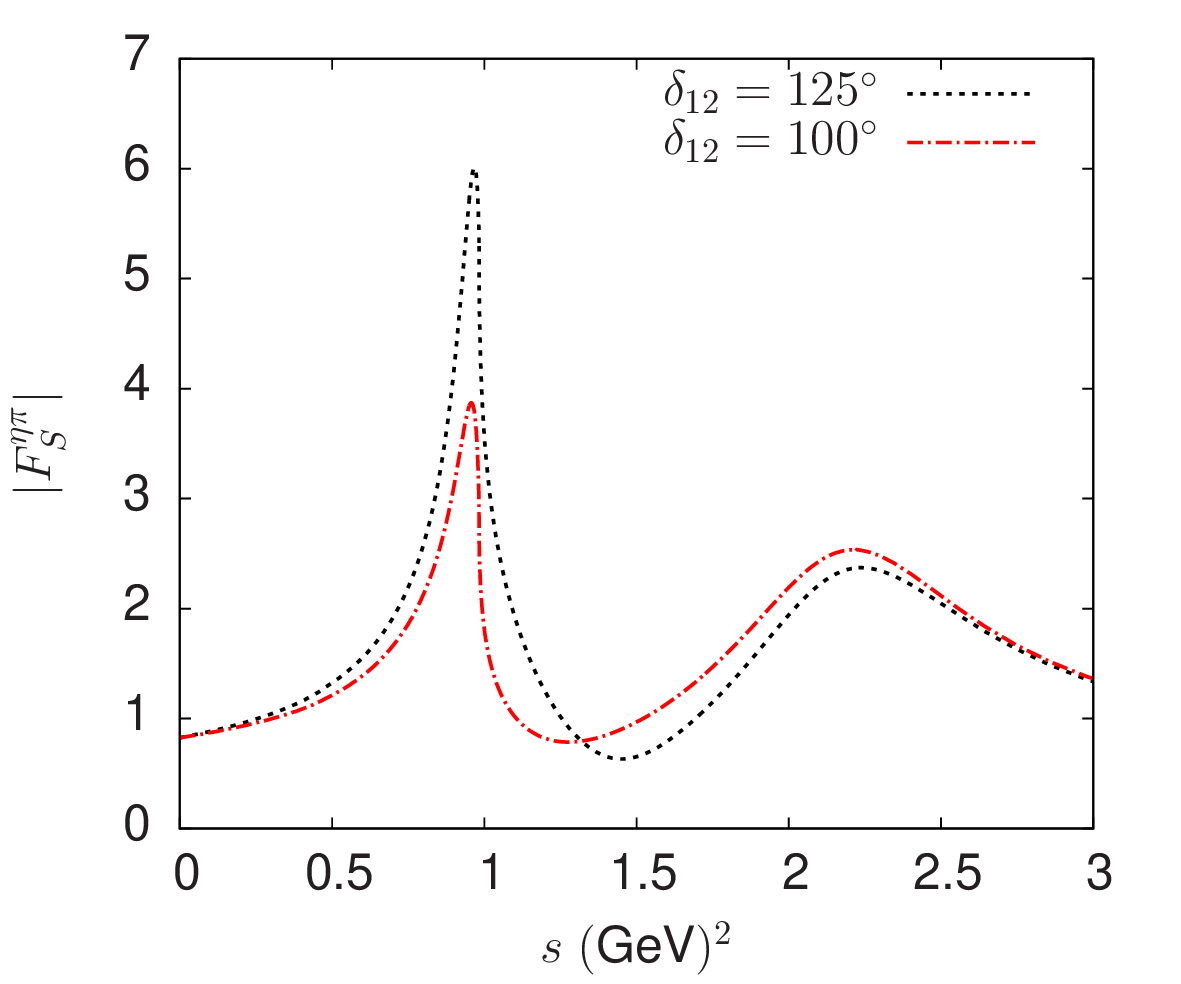}\includegraphics[width=0.499\linewidth]{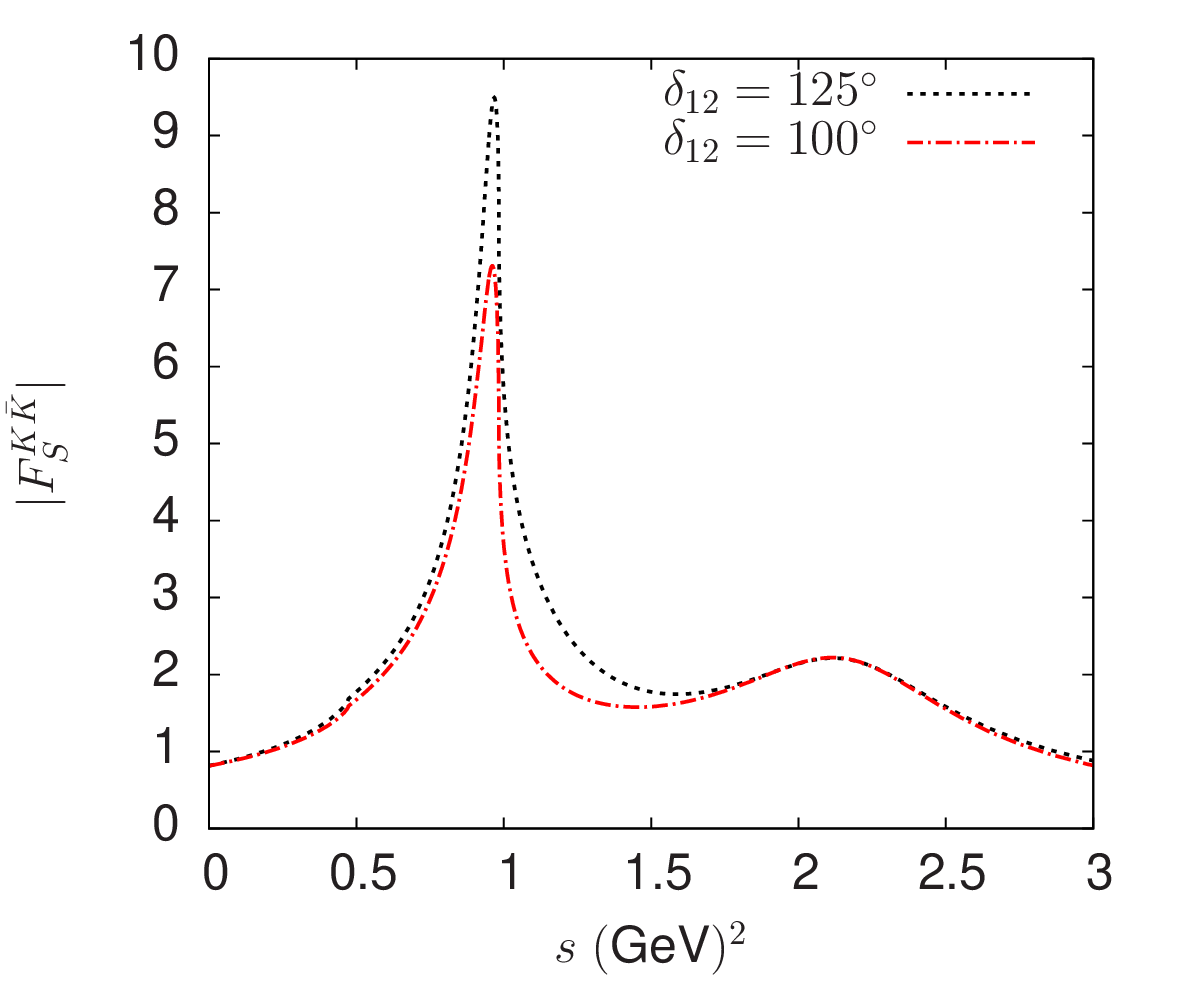}
\caption{\small Absolute values of the form factors $F_S^{\eta\pi}$ (left) and 
$F_S^{K\Kbar}$ (right) computed from our $T$-matrix model, corresponding to
  two input values of the phase $\delta_{12}$.}
\label{fig:FSabsval}
\end{figure}

Finally, fig.~\fig{FSabsval} shows the absolute values of the form factors
$F_S^{\eta\pi}$, $F_S^{K\Kbar}$. The size of the peak associated with
$a_0(980)$ resonance is seen to be sensitive to value of the phase
$\delta_{12}$. We have verified that the associated spectral function agrees
with the one given in ref.~\cite{Guo:2012yt} in the energy range $s < 1.5$
$\hbox{GeV}^2$ when $\delta_{12}\simeq 100\degree$.

\section{Conclusions}
We have proposed a model for the $\eta\pi$ scattering $T$-matrix in the
$S$-wave which satisfies elastic unitarity below the $K\Kbar$ threshold and
two-channel unitarity above. The model is constrained by experimental inputs
on the properties of the two resonances $a_0(980)$, $a_0(1450)$ and by chiral
symmetry at low energy. In the simple $K$-matrix type framework which we have
adopted it is possible to match correctly the two amplitudes
$\eta\pi\to\eta\pi$, $\eta\pi\to K\Kbar$ with the chiral expansion at NLO
while in the case of $K\Kbar\to K\Kbar$, the matching is only approximate (see
sec.~\sect{Tmatrixparam}). Such a $T$-matrix could be realistic in an energy
range $\sqrt{s}\lapprox 1.3$ GeV, where the inelasticity is effectively
dominated by the $K\Kbar$ channel. Formally, however, it is convenient to
extend the model  up to infinite energies such as to allow
for a minimal solution of the associated Muskhelishvili-Omn\`es problem.

A specific prediction of three-flavour ChPT is that the $J=0$ $\eta\pi\to
\eta\pi$ scattering length is very small while the scattering range vanishes
at leading order. The detailed predictions for these quantities at NLO are
very sensitive to the values of the couplings $L_4$, $L_6$ which are $1/N_c$
suppressed. We have used here the values of $L_4$, $L_6$ which are favoured by
lattice QCD simulations. It would be a particularly interesting test of the
chiral expansion, obviously, to have a verification of the $\eta\pi$
scattering length also from lattice QCD.

A supplementary chiral constraint which we have used is associated with the
$\eta\pi$ scalar isovector form factor. We have computed this scalar form
factor from our two-channel $T$-matrix by solving the relevant
Muskhelishvili-Omn\`es integral equations. While this model ignores other
relevant inelastic channels (like $\pi\eta'$) it is nevertheless plausible
that it should be able to describe how the phase of the form factor behaves in
approximately the same energy range where the $T$-matrix is realistic. Above
this point, the model simply serves to interpolate the form factor phase
monotonically towards its known asymptotic value. We find that the small value
of the $\eta\pi$ scalar radius in ChPT at NLO can be understood in this
approach and that this requirement constrains the increase of the sum of
$S$-matrix phases $\delta_{11}+\delta_{22}$ in the $1-2$ GeV energy
region. One should keep in mind the uncertainties on the size of the NNLO
effects on the ChPT side and those from the energy range above 2 GeV on the
dispersive side. The dispersive calculation suggest that the NNLO corrections
to $\braque{r^2}_S^{\eta\pi}$ should tend to increase its size. It would again
be extremely useful to have results from lattice QCD for this quantity.

The computation of the $2\times2$ Omn\`es matrix $\bm{\Omega}$ is a
straightforward extension of the form factor calculation. In principle, the
$\bm{\Omega}$ matrix allows one to treat the $\eta\pi$ rescattering effects
in a unified way, in a number of processes for which recent measurements have
been performed like $\eta'\to\eta\pi\pi$, $\phi\to\eta\pi\gamma$ or
$\gamma\gamma\to\eta\pi$. The consideration of $\eta\pi$ rescattering is also
necessary in the case of the $\eta\to 3\pi$ amplitude in order to account for
$a_0-f_0$ mixing within a dispersive approach.  The $\eta\pi$ scalar form
factor itself appears in the isospin suppressed $\tau\to \eta\pi\nu$
amplitude, along with an electromagnetic induced scalar form factor (and the
vector form factor). This decay mode has not yet been observed but could
possibly be studied at the super-B or future charm-tau factory.

\section*{Acknowledgements}
We would like to thank Jos\'e Antonio Oller for reading the manuscript and
making many useful comments.

This research was supported by Spanish Ministerio de Econom\'ia y
Competitividad and European FEDER funds under contracts FIS2014-51948-C2-1-P, FPA2013-40483-P and FIS2014-57026-REDT, and 
by  the European Community-Research
Infrastructure Integrating Activity "Study of Strongly Integrating
Matter" (acronym HadronPhysics3, Grant Agreement Nr 283286) under the
Seventh Framework Programme of the EU.
M.~A.~acknowledges financial support from the
''Juan de la Cierva'' program (reference 27-13-463B-731) from the Spanish
Government through the Ministerio de Econom\'{\i}a y Competitividad.

\newpage
%%%%%%%%%%%%%%%%%%%%%%%%%%%%%%%%%%%%%%%%%%%%%%%%%%%%%%%%%%%%%%%%%%%%%%
\appendix
\section[The $I=1$ scalar form factors  at NLO]{\boldmath The $I=1$ scalar form factors  at NLO}\lblsec{FFactorsp4}
We consider the two form factors defined in eqs.~\rf{FFactdef}. 
At leading order of the chiral expansion, the form
factors are simply constant,
\be
F_S^{\eta\pi}(0)=\frac{\sqrt6}{3},\quad F_S^{K\Kbar}(0)=1\qquad (LO )\ .
\en 
Computing and adding the next-to-leading order corrections, the form
factors can be written as
\begin{align}
\lbl{Fetapichir}
F_S^{\eta\pi}(s) =F_S^{\eta\pi}(0)\,& \bigg\{1
      + {s\over\fpid} \, \Big[
            4\,L^r_{5}
          + \frac{1}{16\pi^2}\,(- \frac{3}{4})( 1+ L_K)
          \Big]\nonumber\\
&        -\frac{1}{4\fpid}\,(4\,\mkd-3\,s)\,\bar{J}_{KK}(s)
         +\frac{\mpid}{3\fpid}\,\bar{J}_{\pi\eta}(s)\bigg\}\\
\lbl{FKKchir}
F_S^{K\Kbar}(s)= F_S^{K\Kbar}(0)\, & \bigg\{1
       + \frac{s}{\fpid} \, \Big[
            4\,L^r_{5}
          + \frac{1}{16\pi^2} \big(-\frac{1}{4}\big)  \big(  1
          +2 \,R_{\pi\eta} +2\,L_\eta 
          + L_K\big)
          \Big]\nonumber\\
&         + \frac{s}{4\fpid}\, \bar{J}_{KK}(s)\,
        - \frac{1}{6\fpid}\,(4\,m_K^2 - 3\,s)\bar{J}_{\pi\eta}(s) 
\bigg\}~,
\end{align}
where we have introduced the notation
\be
L_P=\log\frac{m_P^2}{\mu^2},\ 
R_{PQ}=\frac{m_P^2\,\log(m_P^2/m_Q^2)}{m_P^2-m_Q^2} ,
\en
and $\bar{J}_{PQ}(s)$ are the loop functions defined to vanish  at $s=0$
(we use the same notation as ref.~\cite{gl85}), 
\begin{align}
\bar{J}_{PQ}(s)& =\frac{s}{16\pi^2}\int_{(m_P+m_Q)^2}^\infty ds'\,
\frac{\sqrt{\lambda_{PQ}(s')}}{(s')^2(s'-s)}\nonumber \\
& =\frac{1}{16\pi^2}\Big[1
+\Big(\frac{\Sigma_{PQ}}{\Delta_{PQ}}
-\frac{\Delta_{PQ}}{s}\Big)\log\frac{m_P}{m_Q}
+\frac{\sqrt{\lambda_{PQ}(s)}}{2s}\,\log\frac
{\Sigma_{PQ}-s+\sqrt{\lambda_{PQ}(s)}}
{\Sigma_{PQ}-s-\sqrt{\lambda_{PQ}(s)}}
\Big]\ 
\end{align}
with
\be
\Sigma_{PQ}=m_P^2+m_Q^2\ ,\quad
\Delta_{PQ}=m_P^2-m_Q^2\ ,\quad
\lambda_{PQ}(s)=s^2-2\Sigma_{PQ}\,s+\Delta_{PQ}^2\ .
\en
The expression for $F_S^{\eta\pi}(0)$ is given by
\begin{align}\lbl{FSetapi(0)}
 F_S^{\eta\pi}(0)=
& {\sqrt6\over 3}\bigg\{
         1
       + {m_K^2\over\fpid} \, \Big[
          - 64\,L^r_{7}
          + 32\,L^r_{6}
          - \frac{32}{3}\,L^r_{5}
          - 16\,L^r_{4}\nonumber\\
&\quad          + \frac{1}{16\pi^2}\,(1 - \frac{2}{9}\,L_\eta 
          + 2\,L_K)
          \Big]\nonumber\\
&       + {\mpid\over\fpid} \, \Big[
            32\,L^r_{8}
          + 64\,L^r_{7}
          + 16\,L^r_{6}
          - \frac{16}{3}\,L^r_{5}
          - 8\,L^r_{4}\nonumber\\
&\quad    + \frac{1}{16\pi^2}\,( - \frac{1}{3}\,R_{\pi\eta} 
          - \frac{5}{18}\,L_\eta 
          - \frac{1}{2}\,L_\pi)
          \Big]\bigg\}
\end{align}
and the expression of $F_S^{K\Kbar}(0)$ reads
\begin{align}\lbl{FKK(0)}
F_S^{K\Kbar}(0)=
&  1
       + \frac{m_K^2}{\fpid} \, \Big[
            16\,(2\,L^r_{8}-L^r_{5})
          + 16\,(2\,L^r_{6}-L^r_{4})
          + \frac{1}{16\pi^2}\big(\,\frac{2}{3}\,R_{\pi\eta} 
           + \frac{10}{9}\,L_\eta\big)
          \Big]\nonumber\\
&       + \frac{m_\pi^2}{\fpid} \, \Big[
            8\,(2\,L^r_{6}-L^r_{4})
          + \frac{1}{16\pi^2}\big(\, - \frac{1}{9}\,L_\eta\big)
          \Big]\ .
\end{align}
\subsection[Remarks on   $F_S^{K\Kbar}(0)$, $F_S^{\eta\pi}(0)$]{\boldmath Remarks on   $F_S^{K\Kbar}(0)$, $F_S^{\eta\pi}(0)$}
The value of $F_S^{K\Kbar}(0)$ can be simply related to the $K^0-K^+$ mass
difference. Indeed, using isospin symmetry, on can express the form
factor $F_S^{K\Kbar}$ as
\be
B_0 F_S^{K\Kbar}(s)=\braque{\frac{\kzero\kzerob-\kplus\kminus}{\sqrt2}\vert
\frac{\ubar{u}-\dbar{d}}{\sqrt2}\vert 0}\ .
\en
Then, writing  the quark masses as
\be
m_u= \hat{m}-\frac{1}{2}\Delta_{du},\quad
m_d= \hat{m}+\frac{1}{2}\Delta_{du}
\en
Feynman-Hellman's theorem yields the following relation,
\be\lbl{FeynmanHellman}
B_0 F^{K\Kbar}_S(0)=\frac{d}{d\Delta_{du}}\left(
M^2_\kzero-M^2_\kplus
\right)\ .
\en
One can easily reproduce eq.~\rf{FKK(0)} using this relation and the 
chiral formula for the mass difference $M^2_\kzero-M^2_\kplus$ from
ref.~\cite{gl85}. Using this formula, one can
also derive an alternative expression for $F_S^{K\Kbar}(0)$,
\be\lbl{altFKK(0)}
F_S^{K\Kbar}(0)= \frac{(\mkd-\mpid)}{(m_s-\hat{m})B_0}\times
\frac{r_2+1}{r+1}
\en
where $r$ is the quark mass ratio $m_s/\hat{m}$ and $r_2= 2\mkd/\mpid-1$ is
the value of this ratio at chiral order $p^2$. The deviation of the value of
$F_S^{K\Kbar}(0)$ from 1 can thus be interpreted as a measure of the size of
the $O(p^4)$ corrections in the chiral expansion of the mass difference
$\mkd-\mpid$. Table~\Table{FSzerovals} below shows that, if one uses the set
of $L_i's$ with large $L_4$, $L_6$, this correction is rather large (of the
order of 40\%). 

We can also perform a verification of the value of $F_S^{\eta\pi}(0)$. Using
the Ward identity in pure QCD,
\be
i\partial_\mu\ubar\gamma^\mu{d}=(m_d-m_u)\,\ubar{d}
\en
we can relate $F^{\eta\pi}_S(0)$ to the value at zero of the $\eta\pi$ vector
form factor $f_+^{\eta\pi}$  (normalized as in  ref.~\cite{Neufeld:1994eg})
when $e^2=0$
\be
F_S^{\eta\pi}(0)= \frac{\sqrt2(\metad-\mpid)}{(m_d-m_u)B_0}
\left.f_+^{\eta\pi}(0)\right\vert_{e^2=0}
\en
Inserting the chiral expansion expressions for $\metad$, $\mpid$ from
ref.~\cite{gl85} and $f_+^{\eta\pi}(0)$ from ref.~\cite{Neufeld:1994eg} one can
recover eq.~\rf{FSetapi(0)}.

The numerical values of $F_S^{\eta\pi}(0)$, $F_S^{K\Kbar}(0)$ are needed as
input in order to solve the integral equations~\rf{FFintequations} for the
scalar form factors. The values at $s=0$ are rather sensitive to the $1/N_c$
suppressed couplings $L_4$, $L_6$ as can be seen from table~\Table{FSzerovals}
below. However, the determination of the scalar radii
$\braque{r^2}_S^{\eta\pi}$, $\braque{r^2}_S^{K\Kbar}$ from the integral
equations depends only on the ratio $F_S^{\eta\pi}(0)/F_S^{K\Kbar}(0)$.  It is
easy to verify that this ratio is independent from $L_4$, $L_6$ at NLO.
%%%%%%
\begin{table}[h]
\centering
\bt{c|ccc}\hline\hline
\        & $O(p^2)$& Small $L_4^r$, $L_6^r$ & Large  $L_4^r$, $L_6^r$ \\
 \hline
$F_S^{\eta\pi}(0)$ & 0.816 & 0.826 & 1.421 \\
$F_S^{K\Kbar}(0)$  &  1    & 0.816 & 1.428  \\ \hline\hline
\et
\caption{\small Numerical values of $F_S^{\eta\pi}(0)$, $F_S^{K\Kbar}(0)$ in
  the chiral expansion at LO and at NLO using two sets of low-energy couplings
(see table~\Table{BE14LECs}).}
\lbltab{FSzerovals}
\end{table}
%%%%%%
\subsection[Expression of $\delta_2$]{\boldmath Expression of $\delta_2$}
We reproduce here the detailed expression (as given in eq. 6.2 of
ref.~\cite{gl85Kl3}) for the term $\delta_2$ which appears in the chiral
expansion of the $K\pi$ scalar radius at order $p^4$ (see eq.~\rf{r2Kpip4})
\be
\delta_2=\frac{-1}{192\pi^2\fpid}\Big[
                                     15\,h_2\Big(\frac{\mpid}{\mkd}\Big)
+\frac{19\,\mkd+3\,\metad}{\mkd+\metad}\,h_2\Big(\frac{\metad}{\mkd}\Big)   
-18      \Big]
\en
with
\be
h_2(x)=\frac{3}{2}\left( \frac{1+x}{1-x}\right)^2
+\frac{3x\,(1+x)}{(1-x)^3}\log(x)\ .
\en
\section[NLO contributions to $I=1$ scattering amplitudes]{\boldmath NLO
  contributions to $I=1$ scattering amplitudes}\lblsec{Decompp4}
We give below the expressions of
the chiral NLO contributions to the one-variable functions associated with the
amplitudes $\eta\pi^+\to \eta\pi^+$, $\eta\pi^+\to \bar{K}^0 K^+$ and 
$\bar{K}^0 K^+\to \bar{K}^0 K^+$.

\subsection[The $\eta\pi^+\to \eta\pi^+$ amplitude]{\boldmath The $\eta\pi^+\to \eta\pi^+$ amplitude}
The $O(p^4)$ part of the amplitude was written in terms of the two
functions $U_0^{11}$, $W_0^{11}$ (eq.~\rf{Decompth}). They can be
expressed as follows
\begin{align}\lbl{etapiU011}
U_0^{11}(s)= & \unsurFQ\bigg\{ 
          (s-\Sigma_{\eta\pi})^2  \, \Big[
            4\,(L^r_{2}+L^r_{3}/3)
          - \frac{3}{8} \frac{1}{16\pi^2}\,( 1+\LK)
          \Big]\nonumber\\
&       + \frac{1}{9}\bar{J}_{\pi\eta}(s)  
           \,m_\pi^4
       +  \frac{1}{24}\,\bar{J}_{KK}(s)  \, 
          (4\,m_K^2-3\,s)^2
\bigg\}\ ,
\end{align}
and
\begin{align}\lbl{etapiW011}%corrected
W_0^{11}(t)= & \unsurFQ\bigg\{ 
         (t-2\,m_\pi^2)\,(t-2\,m_{\eta}^2)  \, \Big[
           4\,(2\,L^r_{1}+L^r_{3}/3)
          - \frac{3}{8} \frac{1}{16\pi^2}\,( 1+ \LK)
          \Big]\nonumber\\
&       + m_\pi^2\,m_{\eta}^2  \, \Big[
          32(-\,L^r_{7}
          + \,L^r_{6}
          - \frac{1}{6}\,L^r_{5}
          - \,L^r_{4})
          + \frac{1}{16\pi^2}\,(\frac{23}{18} + 2\,\LK -
       \frac{2}{9}\,\Leta) 
          \Big]\nonumber\\
&        + m_\pi^4  \, \Big[
            16\,L^r_{8}
          + 32\,L^r_{7}
          + \frac{1}{16\pi^2}\,( - \frac{1}{9} - \frac{2}{9}\,R_{\pi\eta} 
       -\frac{1}{6}\,\LK- \frac{1}{6}\,\Leta - \frac{1}{2}\,\Lpi) 
          \Big]\nonumber\\
&        + t\,\Sigma_{\eta\pi}  \, \Big[
            8\,L^r_{4}
          - \frac{1}{2}  \frac{1}{16\pi^2}\,( 1+\LK)\Big]
        + t\,m_\pi^2  \, \Big[
            \frac{1}{3}\,\frac{1}{16\pi^2}\,\log(\frac{m_K^2}{m_\pi^2})  
          \Big]\nonumber\\
&       - \frac{1}{6}\bar{J}_{\pi\pi}(t)  \, 
          \,m_\pi^2\,(m_\pi^2 - 2\,t)          
       + \frac{1}{54} \bar{J}_{\eta\eta}(t)  \, 
          \,m_\pi^2\,(16\,m_K^2 - 7\,m_\pi^2)
          \nonumber\\
&       - \frac{1}{24} \bar{J}_{KK}(t)  \, 
          \,t\,(8\,m_K^2 - 9\,t)          
\bigg\}\ .
\end{align}
\subsection[The $\eta\pi^+ \to \kzerob\kplus$ amplitude]{\boldmath The $\eta\pi^+ \to \kzerob\kplus$ amplitude}
The three functions involved in the NLO contributions to the amplitude
were denoted as $U_0^{12}$, $W_0^{12}$ and $W_1$. They can be expressed as
\begin{align}
U_0^{12}(s)  =  \sqsixfq\bigg\{ &
       + m_K^4  \,\Big[
          + \frac{16}{3} \,L^r_{8}       
          + \frac{32}{3} \,L^r_{7}
          - \frac{8}{9} \,L^r_{5}
          - \frac{2}{9} \,L^r_{3}
     + \frac{1}{16\pi^2}\,(\frac{1}{72} + \frac{1}{2} \,\Leta -
       \frac{11}{18} \,\LK) 
          \Big]\nonumber\\
&       + m_\pi^2 \,m_K^2  \,\Big[
          - \frac{16}{3} \,L^r_{8}
          - \frac{32}{3} \,L^r_{7}
          + \frac{32}{9} \,L^r_{5}
          + \frac{4}{9} \,L^r_{3}\nonumber\\
& \quad   + \frac{1}{16\pi^2}\,( - \frac{1}{36} - \frac{5}{8} \,\Leta +
       \frac{17}{72} \,\Lpi + \frac{1}{2} \,R_{\eta K} - \frac{13}{9}
       \,R_{\pi K}) 
          \Big]\nonumber\\
&        + m_\pi^4  \,\Big[
          - \frac{2}{9} \,L^r_{3}
          + \frac{1}{16\pi^2}\,(\frac{1}{72} + \frac{1}{8} \,\Leta +
       \frac{4}{9} \,\LK - \frac{41}{72} \,\Lpi 
       - \frac{1}{6}
       \,R_{\eta K} + \frac{11}{18} \,R_{\pi K})
          \Big]\nonumber\\
&        + s \,m_K^2  \,\Big[
          + \frac{1}{16\pi^2}\,( - \frac{1}{12} - \frac{1}{8} \,\Leta +
       \frac{1}{8} \,\LK - \frac{29}{96} \,R_{\eta K} + \frac{67}{96}
       \,R_{\pi K}) 
          \Big]\nonumber\\
&        + s \,m_\pi^2  \,\Big[
          - 2 \,L^r_{5}
          + \frac{1}{16\pi^2}\,(\frac{1}{32} \,\Leta + \frac{1}{16}
       \,\LK + \frac{9}{32} \,\Lpi 
    - \frac{7}{96} \,R_{\eta K} 
    +  \frac{13}{96} \,R_{\pi K}) 
          \Big]\nonumber\\
&        + s^2  \,\Big[
          + \frac{1}{16\pi^2}\,(\frac{1}{16} + \frac{3}{32} \,R_{\eta K} -
       \frac{5}{32} \,R_{\pi K}) 
          \Big] \nonumber \\
&  +(4 \,\mkd-3 \,s)\,\Big[
         \frac{1}{36}\,\bar{J}_{\pi\eta}(s)  \, \,m_\pi^2 
       + \frac{1}{48} \,\bar{J}_{KK}(s) \,s \Big]
\bigg\}~,
\end{align}
and
\begin{align}
W_0^{12}(t)&= \sqsixfq\bigg\{  
         \bar{J}_{K\pi}(t)  \, \Big[
          + \frac{1}{12} \,m_\pi^4
          - \frac{1}{6} \,t \,m_\pi^2
          + \frac{5}{64} \,t^2
          + \frac{1}{12} \,\Delta_{K\pi} \,m_\pi^2
          - \frac{1}{12} \,\Delta_{K\pi} \,t
\nonumber\\
&        + \frac{1}{16} \,\frac{\Delta_{K\pi}^2}{t } \,m_\pi^2
          - \frac{3}{64} \,\Delta_{K\pi}^2
          + \frac{1}{32} \,\frac{\Delta_{K\pi}^3}{t }
          + \frac{1}{48} \,\frac{\Delta_{K\pi}^4}{t^2}
          \Big]
\nonumber\\
&       + \bar{J}_{K\eta}(t)  \, \Big[
          - \frac{5}{36} \,m_\pi^4
          + \frac{1}{6} \,t \,m_\pi^2
          - \frac{3}{64} \,t^2
          - \frac{1}{4} \,\Delta_{K\pi} \,m_\pi^2
          + \frac{1}{6} \,\Delta_{K\pi} \,t
\nonumber\\
&          - \frac{7}{144} \,\frac{\Delta_{K\pi}^2}{t } \,m_\pi^2
          - \frac{43}{576} \,\Delta_{K\pi}^2
          - \frac{19}{288} \,\frac{\Delta_{K\pi}^3}{t }
          + \frac{1}{432} \,\frac{\Delta_{K\pi}^4}{t^2}
          \Big]
\nonumber\\
&      - \frac{1}{48} \,\bar{J}'_{K\pi}(0) \,\frac{\Delta_{K\pi}^4}{t }
       - \frac{1}{432} \,\bar{J}'_{K\eta}(0) \,\frac{\Delta_{K\pi}^4}{t }
\bigg\}~,
\end{align}
and, finally, 
\begin{align}
W_1(t)= \sqsixfq\bigg\{ & 
      \,  t  \, \Big[
          + \frac{1}{3} \,L^r_{3}
          + \frac{1}{16\pi^2}\,( - \frac{1}{48} + \frac{1}{16} \,R_{\eta K} -
       \frac{1}{16} \,R_{\pi K}) 
         \Big] 
\nonumber\\
&        - \frac{1}{64} \, \bar{J}_{K\pi}(t)  \, 
         \frac{\lambda_{K\pi}(t)}{t}
       - \frac{1}{64}\,  \bar{J}_{K\eta}(t)  \, 
         \,\frac{\lambda_{K\eta}(t)}{t}          
\nonumber\\
&        + \frac{1}{16} \,\bar{J}'_{K\pi}(0) \,\Delta_{K\pi}^2
          + \frac{1}{144} \,\bar{J}'_{K\eta}(0) \,\Delta_{K\pi}^2
\bigg\}\ .
\end{align}
\subsection[The amplitude $\kzerob\kplus\to \kzerob\kplus$]{\boldmath The amplitude $\kzerob\kplus\to \kzerob\kplus$}
The $O(p^4)$ contributions to this amplitude involve five functions:
$U_0^{22}$, $U_1$, $V_0$, $V_1$ and $W_0^{22}$ (see
eq.~\rf{Decompth}).  $U_0^{22}$ can be expressed as
\begin{align}
U_0^{22}(s) = &{1\over\fpiq}\bigg\{
            m_K^4 \,\bigg[
           16\,L^r_{8}
          + 32\,L^r_{6}
          - 8\,L^r_{5}
          + \frac{1}{16\pi^2}\,( - \frac{53}{36} 
          +\frac{41}{72}\,\Leta 
    - \frac{3}{4}\,\LK
          - \frac{3}{8}\,\Lpi 
          + \frac{11}{12}\,R_{\pi\eta}) 
          \bigg]\nonumber\\
&       +  s\,m_K^2\,\bigg[
          - 8\,L^r_{4}
          + \frac{1}{16\pi^2}\,(\frac{3}{4} 
          - \frac{1}{16}\,\Leta 
          +\frac{3}{8}\,\LK 
      + \frac{3}{16}\,\Lpi
          -\frac{5}{8}\,R_{\pi\eta}) 
          \bigg]\nonumber\\
&    + s\,m_\pi^2 \,  \bigg[  
          + 2\,L^r_{5}
          + \frac{1}{16\pi^2}\,( - \frac{1}{16}\,\Leta -
   \frac{5}{16}\,\Lpi )
          \bigg]\nonumber\\
&       + (s - 2\,m_K^2 )^2 \,\bigg[
           2\,L^r_{3}
          + 4\,L^r_{2}
          + \frac{1}{16\pi^2}\,( 
          - \frac{1}{24} - \frac{3}{8}\,\Leta 
   +\frac{1}{48}\,\LK - \frac{1}{48}\,\Lpi -
   \frac{3}{8}\,R_{\pi\eta}) 
         \bigg]\nonumber\\
&   +   \frac{1}{24}\,\bar{J}_{\pi\eta}(s)\,(4\,m_K^2 - 3\,s)^2 +
      \frac{1}{16}\,\bar{J}_{KK}(s)\,s^2 
\bigg\}\ .
\end{align}
The function $U_1(s)$ reads
\begin{align}
U_1(s)= &{1\over\fpiq}\bigg\{
         m_K^2 \,\bigg[
           8\,L^r_{4}
          + \frac{1}{16\pi^2}\,( - 1 + \frac{1}{16}\,\Leta -
   \frac{3}{8}\,\LK 
 - \frac{3}{16}\,\Lpi +
   \frac{1}{8}\,R_{\pi\eta}) 
          \bigg]\nonumber\\
&       + m_\pi^2 \,\bigg[
          2\,L^r_{5}
          + \frac{1}{16\pi^2}\,( - \frac{1}{16}\,\Leta -
   \frac{5}{16}\,\Lpi) 
          \bigg]\nonumber\\
&       - \frac{1}{24}\,\bar{J}_{\pi\pi}(s)\,(4\,m_\pi^2 - s) -
   \frac{1}{48}\,\bar{J}_{KK}(s)\,(4\,m_K^2 - s) 
\bigg\}\ .
\end{align}%corrected
Next, the functions $V_0(t)$, $V_1(t)$ read,
\begin{align}\lbl{V0andV1}
   V_0(t) = &{1\over\fpiq}\bigg\{
        ( t- 2\,m_K^2)^2 \,\bigg[
            2\,L^r_{3}
          + 8\,L^r_{1}
          + \frac{1}{16\pi^2}\,( - \frac{29}{48} - \frac{3}{32}\,\Leta
   - \frac{11}{48}\,\LK \nonumber\\
&  - \frac{5}{96}\,\Lpi
   + \frac{3}{16}\,R_{\pi\eta}) \bigg]
   +  \frac{3}{32}\,\bar{J}_{\pi\pi}(t)\,t^2
    +   \frac{1}{288}\,\bar{J}_{\eta\eta}(t)\,(8\,m_K^2 - 9\,t)^2
        \nonumber\\
&    -\frac{1}{48}\,\bar{J}_{\pi\eta}(t)\, 
      (4\,m_K^2 - 3\,t)^2 + \frac{1}{4}\,\bar{J}_{KK}(t)\,t^2
\bigg\}\ ,
\end{align}
\begin{align}
   V_1(t) = &{1\over\fpiq}\bigg\{
      \frac{1}{48}\,\bar{J}_{\pi\pi}(t)\,(4\,m_\pi^2 - t) -
   \frac{1}{12}\,\bar{J}_{KK}(t)\,(4\,m_K^2 - t) 
\bigg\}\ .
\end{align}
Finally, the function $W_0^{22}(u)$ reads,
\begin{align}
   W_0^{22}(u) = &{1\over\fpiq}\bigg\{
        ( 2\,m_K^2 - u)^2 \,\bigg[
           4\,L^r_{2}
          + \frac{1}{16\pi^2}\,( - \frac{7}{24} - \frac{17}{48}\,\LK -
       \frac{1}{48}\,\Lpi) 
          \bigg]\nonumber\\
&    +  \frac{1}{4}\,\bar{J}_{KK}(u)\,(2\,m_K^2 - u)^2
\bigg\}\ .
\end{align}

\section{Asymptotic interpolation}\lblsec{asympinterp}
We describe here the simple ansatz which we used for interpolating the
$S$-matrix phases $\delta_{11}$, $\delta_{22}$ and the inelasticity $\eta$ in
the asymptotic region $s_1 \le s < \infty$. Let $F(s)$ be one of the functions 
$\delta_{11}(s)$, $\delta_{22}(s)$ or $\arccos(\eta(s))$.
We introduce a point
$s_2=s_1+\epsilon$ close to $s_1$ and  we assume that  $F(s_1)$, $F(s_2)$
are given. We denote $s_3=\infty$ and we use the asymptotic conditions
(see~\rf{asycond}) 
\be
\delta_{11}(s_3)= 2\pi,\quad \delta_{22}(s_3)=0,\quad \eta(s_3)=1 \ .
\en
Thus $F(s_3)$ is also known. We introduce a function $u(s)$
\be
u(s)=\dfrac{1}{1+\log\dfrac{s}{s_1}}
\en
which maps the range $[s_1,\infty)$ onto the finite range  $(0,1]$ and define
$F(s)$ through a simple Lagrange polynomial interpolation i.e.
\be
F(s)=\sum_{i=1}^3 F(s_i) \frac{(u(s)-u_j)(u(s)-u_k)}{(u_i-u_j)(u_i-u_k)}
\en
with $u_i\equiv u(s_i)$ and $i, j, k$ is a  cyclic permutation of
$1,2,3$.

\bibliography{essai,uchpt,ffactor}
\bibliographystyle{epj}

\end{document}

%% file: newcommands.tex
\newcommand\TT{\rule{0pt}{2.5ex}}        %top strut// espaces dans les tables
\newcommand{\be}{\begin{equation}}
\newcommand{\en}{\end{equation}}
\newcommand{\bea}{\begin{eqnarray}}
\newcommand{\ena}{\end{eqnarray}}
\newcommand{\lbl}[1]{\label{eq:#1}}
\newcommand{\lbltab}[1]{\label{tab:#1}}

\newcommand{\lblsec}[1]{\label{sec:#1}}
\newcommand{\rf}[1]{(\ref{eq:#1})}
\newcommand{\Table}[1]{\ref{tab:#1}}
\newcommand{\fig}[1]{\ref{fig:#1}}
\newcommand{\sect}[1]{\ref{sec:#1}}
\newcommand{\braque}[1]{{\langle #1 \rangle}}
\newcommand{\outleft}{%
\mathrel{\setbox1=\hbox{$\scriptstyle{out}$}\setbox0=\hbox{$\langle$}\copy0\kern-0.3\wd0\lower1.1\ht0\copy1\kern-0.4\wd1}}

\newcommand{\bc}{\begin{center}}
\newcommand{\ec}{\end{center}}
\newcommand{\bt}{\begin{tabular}}
\newcommand{\et}{\end{tabular}}
\newcommand{\ba}{\begin{array}}
\newcommand{\ea}{\end{array}}
\newcommand{\gapprox}{%
\mathrel{%
\setbox0=\hbox{$>$}\raise0.6ex\copy0\kern-\wd0\lower0.65ex\hbox{$\sim$}}}
\newcommand{\lapprox}{%
\mathrel{%
\setbox0=\hbox{$<$}\raise0.6ex\copy0\kern-\wd0\lower0.65ex\hbox{$\sim$}}}
\renewcommand{\slash}[1]{%
\mathrel{\setbox0=\hbox{$/$}\copy0\kern-\wd0\hbox{$#1$}}}
\newcommand{\im}{{\rm Im\,}}

\newcommand{\metad}{m_\eta^2}

\newcommand{\mpi}{m_\pi}
\newcommand{\mpid}{m_\pi^2}
\newcommand{\fpid}{F_\pi^2}
\newcommand{\mkd}{m_K^2}

\newcommand{\kzero}{{K^0}}
\newcommand{\kplus}{{K^+}}
\newcommand{\kzerob}{{\bar{K}^0}}
\newcommand{\kminus}{{K^-}}
\newcommand{\ubar}{\bar{u}}
\newcommand{\dbar}{\bar{d}}

\newcommand{\Kbar}{\bar{K}}

\makeatletter
\newcommand*\if@single[3]{%
  \setbox0\hbox{${\mathaccent"0362{#1}}^H$}%
  \setbox2\hbox{${\mathaccent"0362{\kern0pt#1}}^H$}%
  \ifdim\ht0=\ht2 #3\else #2\fi
  }
\newcommand*\rel@kern[1]{\kern#1\dimexpr\macc@kerna}
\newcommand*\widebar[1]{\@ifnextchar^{{\wide@bar{#1}{0}}}{\wide@bar{#1}{1}}}
\newcommand*\wide@bar[2]{\if@single{#1}{\wide@bar@{#1}{#2}{1}}{\wide@bar@{#1}{#2}{2}}}
\newcommand*\wide@bar@[3]{%
  \begingroup
  \def\mathaccent##1##2{%
    \if#32 \let\macc@nucleus\first@char \fi
    \setbox\z@\hbox{$\macc@style{\macc@nucleus}_{}$}%
    \setbox\tw@\hbox{$\macc@style{\macc@nucleus}{}_{}$}%
    \dimen@\wd\tw@
    \advance\dimen@-\wd\z@
    \divide\dimen@ 3
    \@tempdima\wd\tw@
    \advance\@tempdima-\scriptspace
    \divide\@tempdima 10
    \advance\dimen@-\@tempdima
    \ifdim\dimen@>\z@ \dimen@0pt\fi
    \rel@kern{0.6}\kern-\dimen@
    \if#31
      \overline{\rel@kern{-0.6}\kern\dimen@\macc@nucleus\rel@kern{0.4}\kern\dimen@}%
      \advance\dimen@0.4\dimexpr\macc@kerna
      \let\final@kern#2%
      \ifdim\dimen@<\z@ \let\final@kern1\fi
      \if\final@kern1 \kern-\dimen@\fi
    \else
      \overline{\rel@kern{-0.6}\kern\dimen@#1}%
    \fi
  }%
  \macc@depth\@ne
  \let\math@bgroup\@empty \let\math@egroup\macc@set@skewchar
  \mathsurround\z@ \frozen@everymath{\mathgroup\macc@group\relax}%
  \macc@set@skewchar\relax
  \let\mathaccentV\macc@nested@a
  \if#31
    \macc@nested@a\relax111{#1}%
  \else
    \def\gobble@till@marker##1\endmarker{}%
    \futurelet\first@char\gobble@till@marker#1\endmarker
    \ifcat\noexpand\first@char A\else
      \def\first@char{}%
    \fi
    \macc@nested@a\relax111{\first@char}%
  \fi
  \endgroup
}
\makeatother